\documentclass[aip,amsmath,amssymb,reprint,nofootinbib,floatfix]{revtex4-1}
\usepackage{graphicx}% Include figure files
\usepackage{dcolumn}% Align table columns on decimal point
\usepackage{bm}% bold math
\usepackage[utf8]{inputenc}
\usepackage[T1]{fontenc}
\usepackage{mathptmx}
\usepackage{comment}

\preprint{AIP/123-QED}
\usepackage{lmodern,graphicx}
\usepackage{times}
\usepackage{mathptmx}
\usepackage{epstopdf}
\usepackage{xcolor}
\usepackage[colorlinks,linkcolor=red,urlcolor=blue,citecolor=red]{hyperref}
\usepackage[normalem]{ulem}
\usepackage{xspace}
\usepackage{float}
\hypersetup{
    colorlinks=true,
    linkcolor=blue,
    filecolor=magenta,      
    urlcolor=cyan,
    citecolor=cyan
}
\usepackage{array}
\newcolumntype{P}[1]{>{\centering\arraybackslash}p{#1}}
\begin{document}

%%%%%%%%%%%%%%%%%%%%%%%%%%%%%%%%%%%%%%%%%%%%%%%%%%%%%%%%%%%%%%%%%%%%
\title[]{Synchronization in the Kuramoto model in presence of stochastic resetting}
% Force line breaks with \\

\author{Mrinal Sarkar}
\affiliation{Department of Physics, Indian Institute of Technology Madras, Chennai 600036, India}
\author{Shamik Gupta}
\affiliation{Department of Physics, Ramakrishna Mission Vivekananda
	Educational and Research Institute, Belur Math, Howrah 711202, India \\Current address: Department of Theoretical Physics, Tata Institute of Fundamental Research, Homi Bhabha Road, Mumbai 400005, India}
\affiliation{Quantitative Life Sciences Section, ICTP–Abdus Salam International Centre for Theoretical Physics, Strada Costiera 11, 34151 Trieste, Italy}
\email{$^1$mrinal@physics.iitm.ac.in;\\$^{2,3}$shamikg1@gmail.com}
\date{\today}

%%%%%%%%%%%%%%%%%%%%%%%%%%%%%%%%%%%%%%%%%%%%%%%%%%%%%%%%%%%%%%%%%%%%%%%%%%%%%%
\begin{abstract}
What happens when the paradigmatic Kuramoto model involving interacting oscillators of distributed natural frequencies and showing spontaneous collective synchronization in the stationary state is subject to random and repeated interruptions of its dynamics with a reset to the initial condition? While resetting to a synchronized state, it may happen  between two successive resets that the system desynchronizes,  which depends on the duration of the random time interval between the two resets. Here, we unveil how such a protocol of stochastic resetting dramatically modifies the phase diagram of the bare model, allowing in particular for the emergence of a synchronized phase even in parameter regimes for which the bare model does not support such a phase. Our results are based on an exact analysis invoking the celebrated Ott-Antonsen ansatz for the case of Lorentzian distribution of natural frequencies, and numerical results for Gaussian frequency distribution. Our work provides a simple protocol to induce global synchrony in the system through stochastic resetting.
\end{abstract}
\maketitle
%\tableofcontents

\begin{quotation}
Studying spontaneous collective synchronization in coupled oscillator systems is a problem of theoretical and practical relevance in nonlinear system studies.  The underlying deterministic dynamics is typically couched in the language of limit-cycle oscillators with distributed natural frequencies that are interacting weakly with one another. With the aim to explore the robustness of the emergent synchronized phase with respect to stochasticity in the dynamics, we unveil here the non-trivial effects induced on subjecting the dynamics to repeated reset to its initial condition at random times.   Our main result concerns existence of the synchronized phase even under conditions in which the bare dynamics does not allow for such a phase.  Our work suggests how a synchronized state may be induced in coupled-oscillator systems through introducing the simple protocol of stochastic resetting, a theme of active research in the area of modern statistical physics,  besides providing a genesis for future work on emergence of intriguing stationary states when purely deterministic dynamics is juxtaposed with the stochastic dynamics of resetting in many-body nonlinear dynamical systems.
\end{quotation}

\section{Introduction}
\label{sec:intro}

Ever since its introduction in the late seventies,  studying collective synchronization in the framework of the celebrated Kuramoto model~\cite{Kuramotobook} has taken centre stage in nonlinear dynamical system studies, with vigorous activities having been pursued by physicists and applied mathematicians;  for a review, see Refs. \cite{Strogatz2000,kuramotormp,Gupta2014,Rodrigues2016,Gupta2018}.  This may be attributed to the ubiquity of the phenomenon of synchronization in nature,  pervading length and time scales of several orders of magnitude,  from yeast cell suspensions, flashing fireflies,  firings of cardiac pacemaker cells, voltage oscillations in Josephson junction arrays,  to animal flocking behavior, pedestrians on footbridges, rhythmic applause in concert halls,  electrical power distribution networks,  and many more~\cite{Pikovskybook, strogatzsync}. This all-prevailing nature of synchronization warrants a theoretical framework to understand how the underlying dynamics may support such an emergent collective phenomenon. The interest in the Kuramoto model is intimately tied to the success it had in explaining analytically how an interacting system of coupled limit-cycle oscillators of distributed natural frequencies may spontaneously evolve to a globally-synchronized stationary state.  In such a state,  a macroscopic number of oscillator phases evolve in time while maintaining a constant phase difference among them.  The original model that involves a bunch of coupled limit-cycle oscillators defined by their phases is premised on a number of assumptions including that there is weak coupling among the oscillators, that everyone is coupled to everyone else with equal strength, and that the interaction between any pair of oscillators depends sinusoidally on the difference of the phases among them~\cite{Kuramotobook}.  The model however allows for straightforward generalizations to factor in a number of features of realistic synchronizing systems, including applying to networks with a given topology,  to have a more general form of inter-oscillator interaction, etc.  For a brief overview of recent developments in the context of Kuramoto dynamics, the reader  is referred to Ref.~\cite{pikovsky2015dynamics}.

A major theoretical result that is also of practical relevance is that for a given distribution of natural frequencies of the oscillators,  the Kuramoto system while starting from generic initial conditions may settle at long times into a globally-synchronized stationary state provided that the strength of interaction between the oscillators exceeds a certain critical value~\cite{Kuramotobook,Strogatz2000,Gupta2018}.  Indeed,  only when the interaction is strong enough that it is able to dominate over the tendencies of individual oscillator phases to rotate independently at their respective frequencies that vary from one oscillator to another.  More precisely,  in terms of a suitably-defined synchronization order parameter that characterizes the presence of global synchrony in the system,  one observes a bifurcation from an unsynchronized to a synchronized phase as the coupling is jacked up beyond a critical value,  with the nature of bifurcation being supercritical or subcritical depending on the nature of the natural frequency distribution~\cite{Strogatz2000,martens2009exact,campa2020phase}.     

The dynamics of the Kuramoto model is inherently deterministic: for the same initial condition on the values of phases of the individual oscillators,  time evolution for a given interval of time results in a unique final configuration of phase values.  A completely different scenario emerges under stochastic dynamics, whereby evolving from the same initial condition for the same interval of time may result in many different final configurations generated with different probabilities.  A paradigmatic and by now a textbook example of stochastic processes is that of Brownian motion~\cite{einstein1956investigations}, originally formulated in the context of the random motion of a mesoscopic particle immersed in a stationary extended medium constituted by microscopic particles and being constantly buffeted by the latter type of particles.  Starting from this rather simple context, the framework of Brownian motion and its many variants have found applications in a wide range of diverse and disparate dynamical scenarios in physics, astronomy, chemistry, biology and mathematics, and even in such interdisciplinary areas as discussing search problems in computer science and fluctuations of stock prices~\cite{majumdar2007brownian,morters2010brownian}. In recent years, stochastic resetting of the Brownian motion has been extensively studied~\cite{r0}.  Here, the Brownian particle while undergoing its intrinsic stochastic dynamics from a given initial condition is reset at random times to the initial condition, following which the dynamics restarts. While a Brownian particle on its own does not have a stationary state in that as time goes on,  it spreads out to larger and larger distances,  a remarkable result of the introduction of resetting is that the particle motion gets effectively confined to a finite interval resulting in a non-trivial stationary state induced by resetting.  Such a confinement emerges not due to application of any confining boundaries but due to random interruptions of the Brownian dynamics through resets to the initial condition that let the particle evolve very far from the initial condition with significantly reduced probability~\cite{r0}. 

In recent times, Ref.~\cite{r0} is widely recognized as having heralded an exciting new beginning in stochastic process studies, and has unleashed a vigorous activity on the theme of stochastic resetting.  The basic paradigm of resetting is that a system which while evolving according to its own intrinsic dynamics resets repeatedly to a given condition (or a set of conditions chosen with given probabilities),  thereby bringing in an intricate interplay of the two time scales characterizing the intrinsic dynamics and the dynamics of resetting. Besides generating fascinating static and dynamic effects arising from this interplay, the dynamics of resetting, which generates non-trivial nonequilibrium stationary states (NESSs)~\cite{r0},  offers a playground to explore uncharted areas in this domain of research.  NESSs are rather unique in that they may exhibit a number of intriguing features that are not observed under equilibrium conditions~\cite{Livi2017}, i.e.,  phase transitions in one dimensional systems with short-range interactions~\cite{mukamel2000phase}. Recent studies of stochastic resetting  include a variety of dynamics: diffusion~\cite{r7, nagar2016diffusion, majumdar2018spectral, den2019properties, r17, masoliver2019telegraphic, ray2020diffusion}, random walks~\cite{montero2016directed,r20}, L\'{e}vy flights~\cite{kusmierz2014first},  Bernoulli trials~\cite{belan2018restart}, discrete-time resets~\cite{coghi2020large}, active motion~\cite{kumar2020active} and transport in cells~\cite{bressloff2020modeling},  search problems~\cite{evans2013optimal, kusmierz2014first, pal2017first, falcon2017localization, chechkin2018random, r12, r21},  RNA polymerases~\cite{r25, r26},  enzymatic reactions~\cite{reuveni2016optimal}, ecology~\cite{boyer2014random,giuggioli2019comparison}, interacting systems such as fluctuating interfaces~\cite{r23,r24}, reaction-diffusion systems~\cite{durang2014statistical},  the Ising~\cite{ising-resetting} model,  exclusion processes~\cite{basu2019symmetric, karthika2020totally}, stochastic thermodynamics~\cite{fuchs2016stochastic}, quantum dynamics~\cite{r4}, etc.

As should be evident from the aforementioned class of work and a survey of the literature on stochastic resetting,  the focus has almost exclusively been on studying the juxtaposition of an intrinsically-stochastic dynamics with stochastic resetting.  Our work departs from this general theme, in that we aim to investigate in this work the issue of what happens when one introduces stochastic resetting into the Kuramoto dynamics, which we have already emphasized to be purely deterministic in nature.  The new features that we may already anticipate in such a mixed dynamics concern stabilization or destabilization of fixed points of the deterministic dynamics. Specifically, in the context of the Kuramoto model, we would like to address: does one continue to observe global synchronization or incoherence of the oscillator phases in the same parameter regime as in the bare model even after introducing stochastic resetting? A minute's thought would convince the reader that the answer may not be in the affirmative.  Indeed, suppose we are in the parameter regime in which the bare Kuramoto dynamics does not exhibit global synchrony in the stationary state.  Nevertheless, repeated interruptions of the dynamics at random times with reset to a synchronized state may allow a stationary synchronized phase in parameter regimes in which it is not observed in the bare dynamics.  A question that naturally arises is: what is the phase diagram of the Kuramoto model dynamics when juxtaposed with stochastic resetting dynamics? 

Besides theoretical relevance,  we may cite the following as motivation for the current study. Sequences of precisely-timed neuronal activity, whereby firing of neurons takes place one after another at precise time intervals, are observed in many brain areas in various species. Classical neural models for such a phenomenon are synfire chain models. A synfire chain is constituted by several pools of spiking neurons connected successively in a feed-forward manner, so that synchronous firing of all neurons in one pool induces a synchronous firing in the next pool, and so on. This eventually leads to a train of synchronous firing patterns in the successive pools.  For an overview,  see Ref.~\cite{abeles1982local}.  In this regard, our model of Kuramoto oscillators with stochastic reset to a synchronized state (a state with $r_0>0$) may be taken to model a common pulse-like drive of neural populations,  resulting in a synfire-chain-like behavior. Namely, all the oscillators on receiving simultaneously a common pulse-like drive are reset at random time intervals to a synchronous state, as may be seen in Fig.~\ref{fig:r_evolution} referred to and discussed later in the text.

\begin{figure*}[!ht]
	\centering
	\includegraphics[scale=0.35]{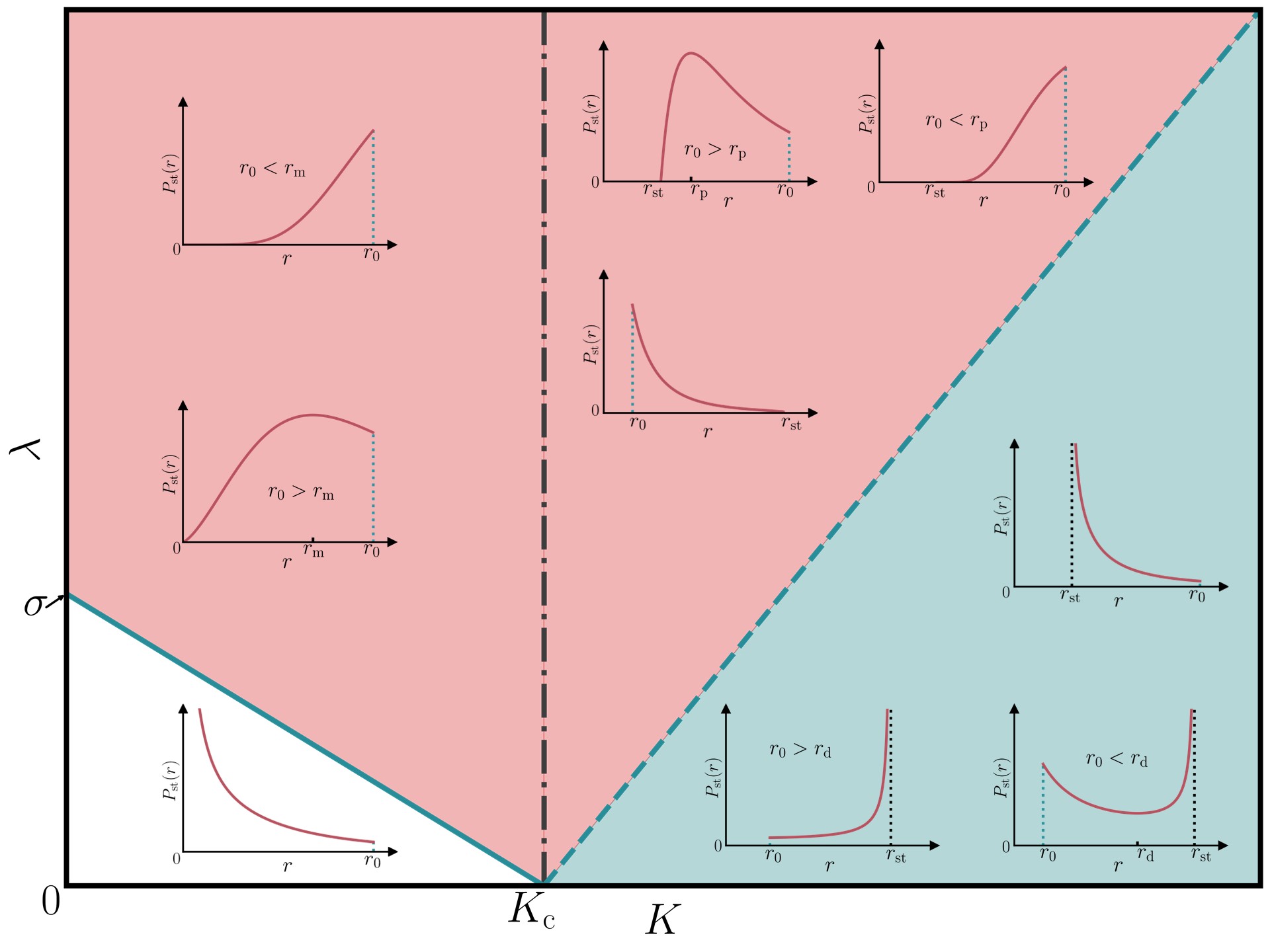}
	\caption{Phase diagram of the Kuramoto model~(\ref{eq:eom0}) subject to stochastic resetting at exponentially-distributed random time intervals and for Lorentzian distribution of natural frequencies with width $\sigma=1.0$, see Eq.~(\ref{eq:Gaussian_lorentzian_g_omega}).  The two axes are the coupling constant $K \ge 0$ and the resetting rate $\lambda \ge 0$. The resetting takes place to states $\{\theta_j\}$ characterized by a given value $r_0$, with $0 < r_0 \le 1$,  of the Kuramoto order parameter~(\ref{eq:order_para_definition}).  In the phase diagram, the whole of the shaded regions corresponds to a synchronized phase, while the white region in the bottom left refers to an unsynchronized phase. This is in the sense that the most probable value of $r_\mathrm{m}$, given by the value of $r$ at which the distribution $P_\mathrm{st}(r)$ peaks, is nonzero in the shaded regions and is zero in the white region. In the diagram, we also show via the different insets the wide spectrum of different behavior exhibited by the stationary-state order parameter distribution $P_{\rm st}(r)$.  The dash-dotted line corresponds to $K=K_c=2\sigma$. The solid line denotes the threshold $\lambda_c$ given by Eq.~(\ref{eq:lambdac}), while the dashed line denotes the threshold $\lambda^\star$ given by Eq.~(\ref{eq:lambdastar}). For details on the behavior of $P_{\rm st}(r)$ as one crosses these lines,   see Sec.~\ref{sec:phase-diagram}. The phase diagram is based on exact results derived in the text in the limit of an infinite number of Kuramoto oscillators. The plots in the various insets have been obtained by choosing appropriate representative values of $r_0$ and $\lambda$ so as to bring out the essential features of the corresponding $P_{\rm st}(r)$. The quantitity $r_{\rm m} \equiv \sqrt{(2\alpha -1)|\Delta|/3}$ corresponds to the $r$-value at which a peak appears in $P_{\rm st}(r)$ for $K < K_c$ and $\lambda > \lambda_{c}$.  The quantity $r_{\rm p}$ (respectively, $r_{\rm d}$) corresponds to the position of a peak (respectively, a dip) that appears in $P_{\rm st}(r)$ for $K > K_c$ and with $\lambda > \lambda^*,  r_0 > r_{\rm {st}}$ (respectively,  with $\lambda < \lambda^*, r_0 < r_{\rm {st}}$, respectively).  We have $r_{\rm p} \equiv \sqrt{\left(2 |\alpha| + 1\right)/3}~ r_{\rm {st}}=r_{\rm d}$ (we use different subscripts for $r_{\rm p}$ and $r_{\rm d}$ to emphasize the fact that although equal,  they are observed in different parameter regimes of $K$ and $\lambda$.).  Here, $\alpha \equiv \lambda/(2 \sigma - K)$, $\Delta \equiv (1- {2\sigma}/{K})$ and $r_{\rm {st}} \equiv \sqrt{1 - K_c/K}$ is the stationary value of $r$, see Subsec.~\ref{subsec:P_st_r0_gt_0} for more details.}  
	\label{fig:phase-diagram}
\end{figure*}

A prominent recent work that in line with our study here considers a deterministic dynamics subject to stochastic resetting is Ref.~\cite{ray2021mitigating}.  In this reference,  the authors study the so-called transient time, namely, the time that it takes to reach a stable equilibrium point in the basin of attraction of a deterministic dynamical system.  It was shown that subjecting the system to stochastic resets leads to a dramatic decrease in the mean transient time.  Our emphasis here is distinctly different in that we want to assess how stochastic resetting affects the phase diagram of a nonlinear dynamical system undergoing deterministic dynamics in time. 

Our main result on the phase diagram of the Kuramoto model subject to stochastic resetting is shown in Fig.~\ref{fig:phase-diagram}. In the phase diagram,  the two axes are the strength of coupling $K \ge 0$ between the oscillators and the resetting rate $\lambda \ge 0$, defined such that in a small time ${\rm d}t$,  the system undergoes a reset with probability $\lambda {\rm d}t$ and evolves according to the deterministic Kuramoto dynamics with the complementary probability $1-\lambda {\rm d}t$.   The diagram, based on exact results derived for a Lorentzian distribution of the natural frequency of the oscillators,  shows parameter regimes in which one observes unsynchronized and synchronized phases.  In particular, the whole of the shaded regions corresponds to a synchronized phase.  We also show in the diagram the wide spectrum of behavior exhibited by the stationary probability distribution of the synchronization order parameter denoted by $r$.  While we will later in the paper discuss the phase diagram in further detail,   we may already point out a remarkable and drastic consequence being induced by resetting.  For $\lambda=0$ (bare Kuramoto dynamics with no resetting),  the system does not show a stationary synchronized phase for values of $K$ smaller than a critical value $K_c$. However, and quite interestingly, one does observe a synchronized phase even in this range of value of $K$ when one switches on resetting in the dynamics by making $\lambda$ non-zero.

The paper is laid out as follows. In Sec.~\ref{sec:model}, we define in detail the Kuramoto model and the dynamics in presence of stochastic resetting. In Sec.~\ref{sec:Lorentzian}, we discuss the case of a Lorentzian distribution for the distribution of natural frequencies of the Kuramoto oscillators. In this section, we first recall in Subsec.~\ref{sec:Ott-Antonsen} a highly-influential exact analytical approach that describes the time evolution of the synchronization order parameter in the bare Kuramoto model. This proves to be an essential ingredient to discuss our analytical results in presence of resetting considered in Subsec.~\ref{sec:with-resetting}, in which we discuss separately the cases $K<K_c$, $K>K_c$, and $K=K_c$.  We discuss the details of the phase diagram~\ref{fig:phase-diagram} in Subsec.~\ref{sec:phase-diagram}. In Sec.~\ref{sec:Gaussian},  for a Gaussian distribution of natural frequencies,  we present in the absence of analytical results several numerical results on the phase diagram.  The paper ends with conclusions in Sec.~\ref{sec:conclusions}. In the Appendix, we discuss the details of numerical simulations reported in the main text.

%%%%%%%%%%%%%%%%%%%%%%%%%%%%%%%%%%%%%%%%%%%%%%%%%%%%%%%%%%%%%%%%%%%%%%%%%%%%%%
\section{Model and dynamics}
\label{sec:model}

The Kuramoto model comprises a collection of $N$ interacting limit-cycle oscillators of distributed natural frequencies that are characterized by their phases, and which are globally coupled through the sine of their phase differences.  Specifically,  the phase $\theta_j (t) \in [0,2\pi)$ of the $j$-th oscillator evolves in time as~\cite{Kuramotobook} 
\begin{align}
	\frac{{\rm d} \theta_{j}}{{\rm d}t}=\omega_{j}+\frac{K}{N}\sum_{k=1}^N \sin(\theta_{k} -\theta_{j}).
	\label{eq:eom0}
\end{align}
Here,  $K \ge 0$ denotes the strength of coupling between the oscillators,  $\omega_j \in (-\infty,\infty)$ is the natural frequency of the $j$-th oscillator,  and the scaling by $N$ of the second term on the right hand side, which may be interpreted as a torque in suitable units,  ensures that this term is well behaved in the limit $N \to \infty$.  The  $\omega_j$s are quenched-disordered random variables distributed according to a common distribution $G(\omega)$ with finite mean $\omega_0>0$ and width $\sigma>0$.  As is usual in studies of the Kuramoto model, we consider $G(\omega)$ to be unimodal, i.e., symmetric about $\omega_0$ and decreasing monotonically and continuously to zero with increasing $|\omega-\omega_0|$. 

Let us introduce the Kuramoto synchronization order parameter~\cite{Kuramotobook,Strogatz2000}
\begin{align}
	r(t) e^{{\rm i} \psi(t)} \equiv \frac{1}{N}  \sum_{j=1}^{N} e^{{\rm i}
		\theta_{j}(t)},
	\label{eq:order_para_definition}
\end{align}
wherein the quantity $r;~0\le r \le 1$ measures the amount of global phase coherence/synchrony present in the system at a given time instant, while $\psi \in [0,2\pi)$ measures the average phase~\cite{Strogatz2000}.  In terms of quantities $r$ and $\psi$, the dynamics~(\ref{eq:eom0}) takes the form
\begin{align}
	\frac{{\rm d} \theta_{j}}{{\rm d}t}=\omega_{j}+Kr(t)\sin(\psi(t) -\theta_{j}),
	\label{eq:eom1}
\end{align}
which makes it evident the mean-field nature of the dynamics: the time evolution of every oscillator is affected by a common mean field of strength $r(t)$ that is generated as a combined effect of all the oscillators. 

We may view the dynamics~(\ref{eq:eom1}) in a frame rotating uniformly with frequency $\omega_0$ with respect to an inertial frame,  which is tantamount to implementing for all $j$ the transformations $\theta_j(t) \to \theta_j(t)-\omega_0 t$ and $\omega_j \to \omega_j - \omega_0$. In the transformed system,  the dynamics~(\ref{eq:eom1}) retains its form, excepting that the $\omega_j$s are now to be considered to be distributed according to the shifted distribution $g(\omega) \equiv G(\omega+\omega_0)$,  thus having zero mean.  We will from now on work in this co-moving frame.  In obtaining our results, we will employ two representative examples of the
frequency distribution, namely, a
Lorentzian and a Gaussian,  given respectively by
\begin{align}
	g(\omega) = \left\{ \begin{array}{lr} \frac{\sigma}{ \pi} \frac{1}{\omega^{2} + \sigma^{2}};&~~(\text{Lorentzian}),
		\\ \frac{1}{\sqrt{2 \pi \sigma^2}} \exp \left(- \omega^{2} / (2 \sigma^{2})\right) ;&~~(\text{Gaussian}),
	\end{array}
	\right.
	\label{eq:Gaussian_lorentzian_g_omega}
\end{align}
with $\sigma$ being identified with the half-width at half-maximum of the Lorentzian distribution, and with the variance of the Gaussian distribution.  

Note that the dynamics~(\ref{eq:eom1}) is deterministic and is moreover non-Hamiltonian.  This latter fact may be understood as follows: Although the torque term may be obtained from a potential ${\cal V}(\{\theta_j\}) \equiv (K/(2N)) \sum_{j,k=1}^N [1 -\cos(\theta_j - \theta_k)]$, a
similar procedure cannot be pursued for the frequency
term. The reason is that an ad hoc potential $\sim - \sum_{j=1}^N \omega_j \theta_j$, which, of course, allows us to obtain the frequency term in the dynamics~(\ref{eq:eom1}) would nevertheless not be periodic in the phases and thus cannot be regarded as a bona fide potential of the system.  Consequently,  the dynamics (\ref{eq:eom1}) cannot be interpreted as a Hamiltonian one describing overdamped motion on a potential landscape:
\begin{align}
\frac{{\rm d}\theta_j}{{\rm d}t}=-\frac{\partial {\cal V}(\{\theta_j\})}{{\rm d}\theta_j},
\label{eq:eom-V}
\end{align}
as is possible if $\omega_j = 0~\forall~j$ when the dynamics does become Hamiltonian.  The dynamics~(\ref{eq:eom1}) relaxes at long times to a stationary state, namely, a state with time-independent values of macroscopic observables,  which in our case are the quantities $r$ and $\psi$.  For $\omega_j = 0~\forall~j$,  the stationary state is an equilibrium one and corresponds to values of $\theta_j$ that minimize the potential ${\cal V}(\{\theta_j\})$.  Otherwise,  the stationary state is a NESS~\cite{Gupta2018}.   

We now briefly summarize the known results for the stationary state of the dynamics~(\ref{eq:eom1}) in the limit $N \to \infty$.  In this limit,  depending on the value of the coupling $K$,  there exist two kinds of qualitatively-different phases characterized by different values of the stationary order parameter $r_{\rm st}\equiv r(t\to \infty)$: an unsynchronized ($r_{\rm st}=0$) phase in the regime $0 \le K \le K_{c}$ and a synchronized ($0 < r_{\rm st} \leq 1$) phase in the regime $K > K_{c}$,  with $K_c$,  the critical coupling strength, given by $K_{c} = 2/(\pi g(0))$.  The latter result is obtained on a basis of an ingenious self-consistent analysis of the dynamics proposed by Kuramoto himself~\cite{Kuramotobook}. The dynamics (\ref{eq:eom1}) exhibits a supercritical bifurcation between the two phases as one tunes $K$ across $K_{c}$~\cite{Kuramotobook,Strogatz2000,Gupta2018}. Considering the behavior of $r_{\rm st}$ as a function of $K$,  one may draw a parallel of the bifurcation behavior with that of a continuous phase transition in statistical physics~\cite{Gupta2018,Livi2017}. 

\begin{figure}[!ht]
	\centering
	\includegraphics[scale=0.35]{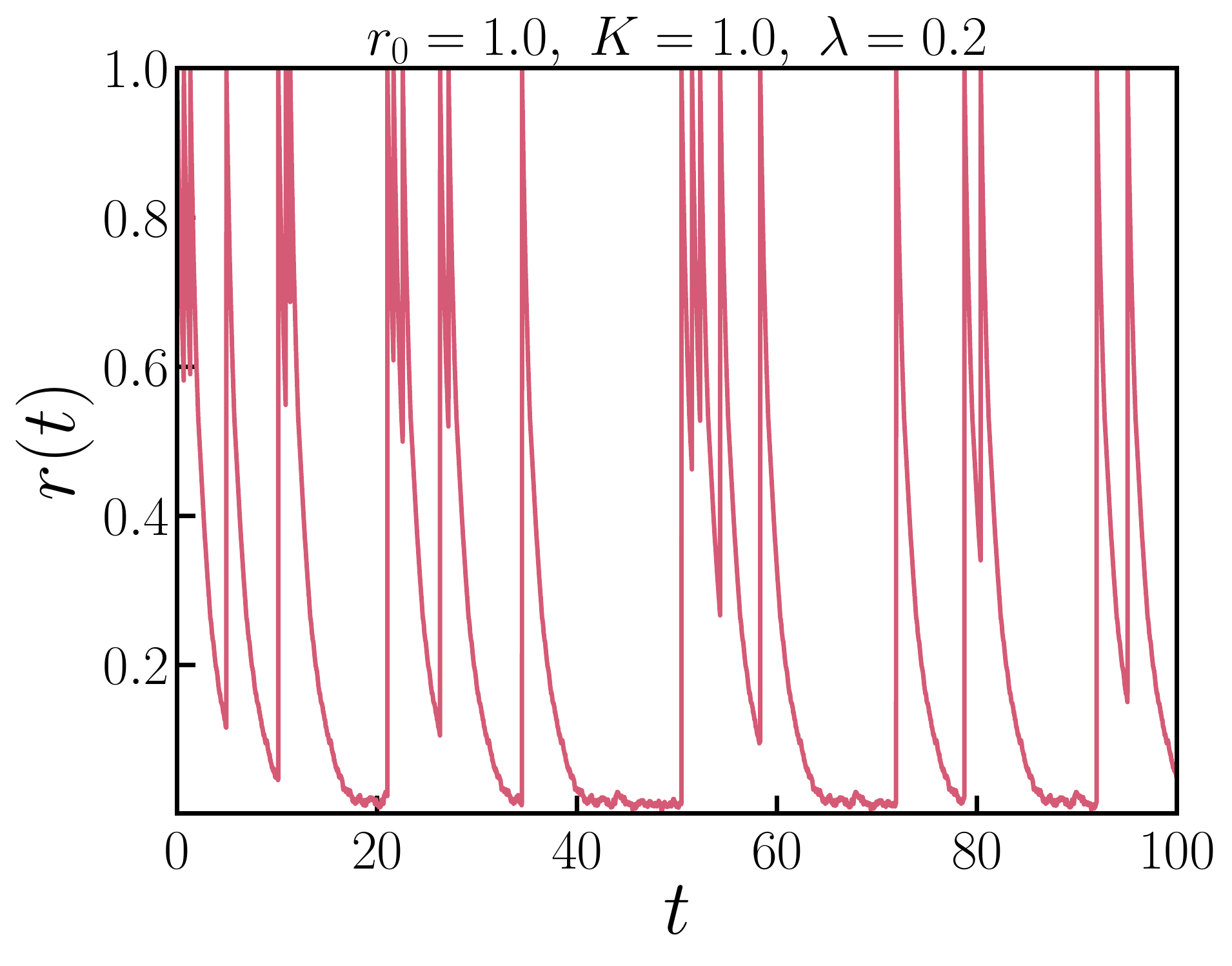}
	\caption{Shown is the variation of order parameter $r$ as a function of time in a single realization of the Kuramoto dynamics with resetting, with number of oscillators $N=10^4$ and for the choice of parameters: $r_0=1$, $K=1.0$, and $\lambda = 0.2$. The data correspond to a Lorentzian frequency distribution with width $\sigma=1.0$, Eq.~(\ref{eq:Gaussian_lorentzian_g_omega}). The detailed numerical procedure to generate the data is provided in the Appendix.}  
	\label{fig:r_evolution}
\end{figure}

We now discuss how resetting may be introduced into the dynamics~(\ref{eq:eom1}).  To this end,  the dynamics starting with a given initial state $\{\theta_j(0)\}$ is interrupted at random times at which the system is instantaneously reset to the initial condition and following which the dynamics restarts, see Appendix for details of numerical implementation. Thus,  a typical realization of the dynamics for a given initial state would consist of deterministic evolution governed by Eq.~(\ref{eq:eom1}) interspersed with instantaneous resets at random times to the initial state;  between any two successive resetting events, the oscillator phases evolve deterministically following~(\ref{eq:eom1}). The random times at which a reset happens follow a Poisson point process with rate $\lambda$.  This implies that the random variable $\tau$,   denoting the interval between two successive resets,  is distributed according to an exponential distribution
\begin{align}
p(\tau)=\lambda e^{-\lambda \tau};~~\lambda \ge 0,~\tau [0,\infty),
\label{eq:exponential}
\end{align}
which may equivalently be viewed as implementing in every infinitesimal time interval between times $t$ and $t+{\rm d}t$ the state $\{\theta_j(t)\}$ to either evolve according to the dynamics~(\ref{eq:eom1}) with probability $1-\lambda {\rm d}t$ or be reset to the initial state $\{\theta_j(0)\}$ with probability $\lambda {\rm d}t$.  Here $\lambda$ is the resetting rate,  and the quantity $1/\lambda$ denotes the average time between two successive resets.  Evidently then, on setting $\lambda$ to zero, one recovers the bare Kuramoto dynamics~(\ref{eq:eom1}) with no resets.  We consider in this work class of initial states characterized by a given value $r_0\equiv r(t=0)$,  with $0 \le r_0 \le 1$, of the macroscopic order parameter $r$.  We may at this point list down the various parameters that characterize the Kuramoto dynamics in presence of stochastic resetting: the coupling strength $K$,  the width $\sigma$ of the frequency distribution,  the reset value $r_0$ of the order parameter, the quantity $\lambda$ that sets the time scale for reset. A representative variation of $r$ as a function of time in a single realization of the dynamics and for chosen values of the various parameters is shown in Fig.~\ref{fig:r_evolution}. The data correspond to a Lorentzian frequency distribution and have been generated by following the numerical procedure detailed in the Appendix.

We expect nontrivial stationary states in the Kuramoto model in presence of resetting. From the bare dynamics~(\ref{eq:eom1}), we may observe that $K$ has the dimension of inverse time. Moreover, the parameter $K$ sets the scale over which effects of interactions between the oscillators come into play in determining the evolution of the $\theta_j$s. Then, in presence of resetting,  the two competing time scales corresponding to the bare Kuramoto dynamics and the one due to stochastic resets would be set respectively by the quantities $K$ and $\lambda$.  Choosing $K$ to be smaller than $K_c$,  and for comparable values of $\lambda$,  the bare dynamics would attempt to induce in the dynamics a tendency in $r$ to attain the value $r_{\rm st}(K<K_c)=0$. The resetting dynamics would on the other hand induce a competing tendency in $r$ to attain values around $r_0$, for values of $r_0$ larger than zero.  Hence, on the basis of the foregoing, we may already anticipate that depending on the values of $K$, $\lambda$ and $r_0$,  a synchronized phase may emerge for values of $K$ for which the bare dynamics does not allow for any such phase.        

In the aforementioned backdrop, it is pertinent to ask: How does the inclusion
of the stochastic resetting protocol modify the stationary-state phase diagram of the bare Kuramoto model discussed above,  which is now to be considered in the two-dimensional $(K,\lambda)$-plane for a given value of $r_0$? Do new
phases emerge and can one characterize analytically the different phases and obtain the boundaries between them? An analytical study does not certainly seem trivial in light of the many-body and non-Hamiltonian nature of the dynamics~(\ref{eq:eom1}).  We are tasked with answering the aforementioned questions in the rest of the paper.

%%%%%%%%%%%%%%%%%%%%%%%%%%%%%%%%%%%%%%%%%%%%%%%%%%%%%%%%%%%%%%%%%%%%%%%%%%%%%%
\section{Lorentzian frequency distribution: Exact results}
\label{sec:Lorentzian}

\subsection{Analysis in absence of resetting: The Ott-Antonsen ansatz}
\label{sec:Ott-Antonsen}
Before we proceed to embark on an analysis of the dynamics~(\ref{eq:eom1}) in presence of stochastic resetting,  we will recall in this section an approach owing to Ott and Antonsen. This approach allows to analyze the Kuramoto dynamics in absence of resetting, which will prove useful in our analysis later in the paper.  Central to this approach is the so-called Ott-Antonsen (OA) ansatz that is based on the following premises: In the limit $N \to \infty$, the state of the oscillator system~(\ref{eq:eom1}) at time $t$ may be characterized by the time-dependent single-oscillator
distribution function $F(\theta,\omega,t)$, defined such that $F(\theta,\omega,t){\rm
	d}\theta$ gives the fraction of oscillators with frequency $\omega$ that
have their phase in $[\theta,\theta+{\rm d}\theta]$ at time $t$. The
function $F(\theta,\omega,t)$ is $2\pi$-periodic in $\theta$:
\begin{align}
F(\theta+2\pi,\omega,t)=F(\theta,\omega,t), 
\end{align}
and is normalized as 
\begin{align}
	\int_0^{2\pi}{\rm d}\theta~F(\theta,\omega,t)=g(\omega)~\forall~\omega, ~t.
	\label{eq:f-normalization}
\end{align} 
For the case of the Lorentzian distribution of oscillator frequencies given in Eq.~(\ref{eq:Gaussian_lorentzian_g_omega}),  the Ott-Antonsen ansatz stipulates that in the space ${\cal D}$ of all possible distribution functions $F(\theta, \omega, t)$, there exists a particular class defined on and remaining confined to a manifold ${\cal M} \in {\cal D}$ under the time evolution dictated by the Kuramoto dynamics~\cite{ott2008low,ott2009long}.  Implementing the ansatz yields a single first-order ordinary differential equation for the evolution of the synchronization order parameter $r(t)$, which may be solved exactly to obtain how $r(t)$ evolves in time starting from an arbitrary initial condition $r(0)$.  The power and the usefulness of the ansatz and its wide applicability across phase oscillator systems stem from its remarkable ability to capture quantitatively through this single equation all, and not just some, of the order parameter attractors and bifurcations of the dynamics (\ref{eq:eom1}) (which may be obtained by performing numerical integration of the $N$ coupled non-linear equations contained in Eq.~(\ref{eq:eom1}) for $N \gg 1$ and a Lorentzian $g(\omega)$, and then evaluating $r(t)$ in numerics).  The approach has ushered in a slew of interesting and insightful theoretical work by physicists and applied mathematicians working  in the field of nonlinear dynamics.  A selection of recent publications is given by Refs.~\cite{ott0,ott11,ott10,ott9,ott8,ott7-5,chandrasekar2020kuramoto,ott6,ott5,ott4,ott3,ott2,ott-shamik,ott1-again,ott1,ott-m1}.  For a recent review on the OA ansatz, the reader is referred to Ref.~\cite{OAreview}.

To delve further into the OA approach,  let us write down the time evolution equation for the distribution function $F(\theta,\omega,t)$. To this end, let us first note that the dynamics~(\ref{eq:eom1}) evidently conserves in time the number of
oscillators with a given natural frequency,  and consequently,  the function  
$F(\theta,\omega,t)$ evolves in time according to the continuity equation
\begin{align}
	\frac{\partial F}{\partial t}+\frac{\partial}{\partial \theta}\left(F\frac{{\rm
		d}\theta}{{\rm d}t}\right)=0,
	\label{eq:continuity_eqn}
\end{align}
wherein the quantity ${\rm d}\theta/{\rm d}t$ is obtained from Eq.~ (\ref{eq:eom1}) as
\begin{align}
	\frac{{\rm d} \theta}{{\rm d}t} =  \omega + \frac{Kr(t)}{2{\rm i}}[ e^{{\rm i} (\psi(t) -\theta)} -  e^{-{\rm i} (\psi(t) -\theta)}],
	\label{eq:vel}
\end{align}
where 
\begin{align}
	r(t) e^{{\rm i} \psi(t)} \equiv \int {\rm d}\theta {\rm d}\omega~e^{{\rm i}\theta}F(\theta,\omega,t)
	\label{eq:rf-definition}
\end{align}
is obtained as the $N \to \infty$-generalization of Eq.~(\ref{eq:order_para_definition}).
On substituting Eq.~(\ref{eq:vel}) in Eq.~(\ref{eq:continuity_eqn}), we obtain
\begin{align}
	&\frac{\partial F(\theta,\omega,t)}{\partial t}+\omega\frac{\partial
		F(\theta,\omega,t)}{\partial
		\theta}+\frac{K}{2{\rm i}}\frac{\partial }{\partial
		\theta}\Big[r(t)\Big(e^{{\rm i}(\psi(t) -\theta) }\nonumber \\
	&\qquad - e^{-{\rm i} (\psi(t) -\theta)}\Big
	)F(\theta,\omega,t)\Big]=0.
	\label{eq:kinetic-equation}
\end{align}

Being $2\pi$-periodic in $\theta$, we may apply a Fourier expansion of $F$ in $\theta$ thus:
\begin{align}
	F(\theta,\omega,t)=\frac{g(\omega)}{2\pi}\Big[1+\sum_{n=1}^\infty
	\Big(\widetilde{F}_n(\omega,t)e^{{\rm i}n\theta}+[\widetilde{F}_n(\omega,t)]^\ast
	e^{-{\rm i}n\theta}\Big)\Big],
	\label{eq:f-expansion}
\end{align}
where $\widetilde{F}_n(\omega,t)$ is the $n$-th Fourier coefficient, and $\ast$ stands for complex conjugation. Using $\int_0^{2\pi}{\rm
	d}\theta~e^{{\rm i}n\theta}=2\pi \delta_{n,0}$,  it is easily checked that the above
expansion indeed satisfies Eq.~(\ref{eq:f-normalization}).

We now proceed to employ the OA ansatz~\cite{ott2008low,ott2009long} that
allows to derive from Eq.~(\ref{eq:kinetic-equation}) an ordinary differential equation for the order parameter $r$.  As is usual with OA ansatz implementation, we will make the specific choice of a Lorentzian $g(\omega)$ as given in Eq.~(\ref{eq:Gaussian_lorentzian_g_omega}). 
The ansatz considers in the expansion~(\ref{eq:f-expansion}) a restricted class of Fourier coefficients given
by~\cite{ott2008low,ott2009long}
\begin{align}
	\widetilde{F}_n(\omega,t)=[z(\omega,t)]^n,
	\label{eq:OA}
\end{align}
with $z(\omega,t)$ an arbitrary function with the restriction
$|z(\omega,t)|<1$ that makes the infinite series in
Eq.~(\ref{eq:f-expansion}) a convergent one. In implementing the OA
ansatz, it is also assumed that $z(\omega,t)$ may be
analytically continued to the whole of the complex-$\omega$ plane, that it has no singularities in the lower-half complex-$\omega$ plane, and that
$|z(\omega,t)|\to 0$ as ${\rm Im}(\omega) \to -\infty$~\cite{ott2008low,ott2009long}. 
Using Eqs.~(\ref{eq:f-expansion}) and~(\ref{eq:OA}) in
Eq.~(\ref{eq:rf-definition}), one arrives at the result
\begin{align}
	r(t)e^{{\rm i} \psi(t)}=\int_{-\infty}^\infty {\rm
		d}\omega~g(\omega)z^\ast(\omega,t). 
	\label{eq:rf-OA}
\end{align}
On substituting Eqs.~(\ref{eq:f-expansion}),~(\ref{eq:OA}),
and~(\ref{eq:rf-OA}) in Eq.~(\ref{eq:kinetic-equation}) and on collecting and equating the
coefficient of $e^{{\rm i}n\theta}$ to zero,  one obtains
\begin{align}
	&\frac{\partial z(\omega,t)}{\partial t}+{\rm i}\omega
	z(\omega,t)+\frac{Kr(t)}{2}\Big[
	z^2(\omega,t)e^{{\rm i} \psi(t)}-e^{-{\rm i} \psi(t)}\Big]=0. \nonumber \\
	\label{eq:lambda-differential-equation}
\end{align}

For the Lorentzian $g(\omega)$, see Eq.~(\ref{eq:Gaussian_lorentzian_g_omega}), one may
evaluate $r(t)$ by using Eq.~(\ref{eq:rf-OA}) to get
\begin{eqnarray}
	r(t)e^{{\rm i} \psi(t)} &=&\frac{1}{2i\pi}\oint_C {\rm
		d}\omega~z^\ast(\omega,t)\left[\frac{1}{\omega-{\rm i}\sigma}-\frac{1}{\omega+{\rm i}\sigma}\right]\nonumber
	\\
	&=&z^\ast(-{\rm i}\sigma,t),
	\label{eq:rf-integral}
\end{eqnarray}
where the contour $C$ consists of the real-$\omega$ axis closed by a
large semicircle in the lower-half complex-$\omega$ plane on which the
integral in Eq.~(\ref{eq:rf-integral}) gives zero contribution in view of
$|z(\omega,t)|\to 0$ as ${\rm Im}(\omega) \to -\infty$. The second
equality in Eq.~(\ref{eq:rf-integral}) is obtained by applying the
residue theorem to evaluate the complex integral over the contour $C$.
Using Eqs.~(\ref{eq:lambda-differential-equation})
and~(\ref{eq:rf-integral}), we finally obtain the OA equation for the time
evolution of the synchronization order parameter as the single ordinary differential equation~\cite{ott2008low}
\begin{align}
	&\frac{{\rm d} r(t)}{{\rm d}
		t}+f(r)=0;~~f(r) \equiv  \left[\sigma- \frac{K}{2}\right]r+ \frac{K}{2}r^3,
	\label{eq:rt-differential-equation}
\end{align}
while the average phase $\psi(t)$ evolves as  ${{\rm d} \psi}/{{\rm d}t} = 0$,  implying thereby that under the OA ansatz,  $\psi$ does not evolve from its initial value. The solution to Eq.~(\ref{eq:rt-differential-equation}) requires as an initial condition only the value of $r(t)$ at $t=0$.

The dynamics~(\ref{eq:rt-differential-equation}) may be put in the form of an overdamped dynamics on a potential landscape:
\begin{align}
\frac{{\rm d}r}{{\rm d}t}=-V'(r);~~V(r)=\left(\sigma-\frac{K}{2}\right)\frac{r^2}{2}+\frac{Kr^4}{8},
\label{eq:r-V}
\end{align}
with prime denoting derivative with respect to $r$. Note that $r=0$ is a fixed point of the dynamics for all values of $K$ and $\sigma$. The stationary values of $r$ are those fixed points of the dynamics~(\ref{eq:r-V}) that correspond to the minima of the potential $V(r)$;  the stationary values are by definition linearly stable under the dynamics~(\ref{eq:r-V})~\cite{strogatz-book}. One obtains 
\begin{equation}
r_{\rm st}=\left\{ 
\begin{array}{ll}
               0 & \mbox{for~$K< 2\sigma$}, \\
               \sqrt{1-\frac{2\sigma}{K}} & \mbox{for~$K > 2\sigma$}. 
               \end{array}
        \right. 
\label{eq:rst-Kuramoto}
\end{equation} 
We thus obtain the critical $K$ at which $r_{\rm st}$ goes over from a zero to a non-zero value as equal to $2\sigma$,  which coincides with the bifurcation point $K_c=2/(\pi g(0))$, with $g(0)$ given by the Lorentzian in Eq.~(\ref{eq:Gaussian_lorentzian_g_omega}), that is known for the dynamics~(\ref{eq:eom1}) from the self-consistent analysis of Kuramoto.  Note that we will in this work use the notation $r_{\rm st}$ to denote the stationary value of $r$ in absence of resetting, while the mean stationary value in its presence will be denoted by $r_{\rm st}^{(\lambda)}$, see later.

We emphasize that Eq.~(\ref{eq:rt-differential-equation}) and its stationary-state solution given by Eq.~(\ref{eq:rst-Kuramoto}) hold only for a Lorentzian $g(\omega)$ and, more importantly,  on assuming the validity of the OA ansatz~(\ref{eq:OA}).  In the absence of this ansatz,  a differential equation describing the dynamics of the order parameter $r(t)$ under the Kuramoto dynamics~(\ref{eq:eom1}) is not known, and in fact, there is a priori no reason why the dynamics given by the ansatz should coincide with the one obtained, e.g., by numerically integrating the equations of motion~(\ref{eq:eom1}).  However, as mentioned in the leading paragraph of this subsection,  it has indeed been demonstrated that the two analyses do coincide, which hints at the attracting property of the OA manifold~\cite{ott2008low,ott2009long}.      

\subsection{Analysis in presence of resetting}
\label{sec:with-resetting}

Resetting promotes the status of the order parameter $r(t)$ to that of a random variable. It therefore behoves us to define a probability density in order to characterize the very many values that $r$ may take at time $t>0$ in different realizations of the resetting dynamics, while starting from the same initial value $r_0$ and resetting at exponentially-distributed random time intervals to $r_0$.  We will choose $r_0$ to be any value on the OA manifold; For details on how to implement such a choice in performing numerical simulations, see Appendix. To this end,  define $P(r,t)\equiv P(r,t|r_{0}, 0;\lambda)$ as the probability density that the order parameter has the value $r$ at time $t$, given that it had the value $r_{0}$ at time $t=0$. This probability density is normalized as 
\begin{align}
	\int_{0}^{1} {\rm d}r~P(r,t) = 1~\forall~ t.
	\label{eq:P_normalization}
\end{align}
The time evolution of $P(r,t)$ is given by the Master equation
\begin{equation}
\frac{\partial P(r,t) }{\partial t} = - \frac{\partial}{\partial r} \left(P(r,t)\frac{{\rm d}r}{{\rm d}t}\right) - \lambda  P(r,t) + \lambda \delta (r-r_{0}),	
\label{eq:ME-1}
\end{equation}

with the initial condition $P(r,0)=\delta (r-r_{0})$.  The first term on the right hand side (rhs) of Eq.~(\ref{eq:ME-1}) represents the contribution due to the bare Kuramoto dynamics in absence of resetting given by Eq.~(\ref{eq:eom1}), and in fact, setting $\lambda$ to zero in the equation reduces it to the continuity equation that expresses the conservation of probability under the deterministic dynamics~(\ref{eq:eom1}). The second and third terms on the rhs owe their presence to the process of resetting:  The second term represents a loss in probability due to resetting from $r\ne r_0$ to $r_{0}$, while the third term denotes the corresponding gain in probability at $r=r_{0}$. 

In order to proceed,  we need to use in Eq.~(\ref{eq:ME-1}) the dynamics of $r$ in time contained in the term ${\rm d}r/{\rm d}t$. As we have already commented,  this is known only under the OA ansatz applicable for a Lorentzian $g(\omega)$, which is what we invoke to make headway into the problem. To this end, using Eq.~(\ref{eq:rt-differential-equation}) in Eq.~(\ref{eq:ME-1}) yields 
\begin{eqnarray}
	\frac{\partial P(r,t) }{\partial t} = \frac{\partial}{\partial r} [f(r) P(r,t)] - \lambda  P(r,t) + \lambda \delta (r-r_{0}).
	\label{eq:ME-2}
\end{eqnarray}

Our primary interest is to ascertain the effects of resetting on the stationary state.  As mentioned in the Introduction and as has been amply emphasized in the literature,  introducing resetting into a dynamical scenario results in the modified dynamics relaxing to an NESS. The latter in our case is characterized by the stationary probability $P_{\rm {st}}(r)$, which from Eq.~(\ref{eq:ME-2}) may be seen to satisfy
\begin{eqnarray}
\frac{\rm d}{{\rm d}r} \left[f(r)P_{\rm {st}}(r)\right] - \lambda P_{\rm {st}}(r) =   - \lambda \delta (r-r_{0}).
\label{eq:ME-3-st-0}
\end{eqnarray}

\subsubsection{The case $r_0=0$}
In this case,  one may easily check that 
\begin{align}
P_{\rm st}(r)=\delta(r),
\label{eq:ME-st-r0-0}
\end{align}
with the understanding that $r \ge 0$, satisfies Eq.~(\ref{eq:ME-3-st-0}); while on direct substitution, the second term on the left hand side cancels with the term on the right hand side, the first term on the left hand side evaluates to zero: $({\rm d}/{\rm d}r)\left[f(r)\delta(r) \right]= ({\rm d}/{\rm d}r)\left[f(0)\delta(r) \right]=({\rm d}/{\rm d}r)[0~\delta(r)]=0$. Here, we have used the fact that $f(0)=0$. Note that Eq.~(\ref{eq:ME-st-r0-0}) is the stationary state distribution for all values of $K$ and $\lambda$. This result may be understood as follow: For the Kuramoto model, $r_0=0$ is a fixed point of the OA-dynamics for all values of $K$ and $\sigma$, and so when the bare dynamics is initiated with $r_0=0$,  there is no evolution of $r(t)$. The situation remains the same on introducing resetting to $r_0=0$, and consequently, the $r$-distribution is of the form (\ref{eq:ME-st-r0-0}).

\subsubsection{The case $0 < r_0 \le 1$}
\label{subsec:P_st_r0_gt_0}
We will now consider the case $0 < r_0 \le 1$. In this case, we first rewrite Eq.~(\ref{eq:ME-3-st-0}) as
\begin{align}
\frac{\rm d}{{\rm d}r} P_{\rm {st}}(r) + \left[ \frac{f'(r) - \lambda}{f(r)}\right] P_{\rm {st}}(r) = - \frac{\lambda}{f(r)} \delta (r-r_{0}).
	\label{eq:ME-3-st}
\end{align}
The presence of the delta function in Eq.~(\ref{eq:ME-3-st}) immediately implies a discontinuity in $P_{\rm {st}}(r)$ at $r=r_{0}$. To obtain the discontinuity, we integrate Eq.~(\ref{eq:ME-3-st}) between $[r_{0} - \epsilon, r_{0} + \epsilon]$ with $0<\epsilon \ll 1$, to arrive at
\begin{equation}
	P_{\rm {st}}(r_{0} + \epsilon) - P_{\rm {st}}(r_{0} - \epsilon) =  - \frac{\lambda}{f(r_{0})}.
	\label{eq:P_discont}
\end{equation}

For $r<r_0$ and $r>r_0$,  Eq.~(\ref{eq:ME-3-st}) is a first-order homogeneous ordinary differential equation with constant coefficients, which can be easily solved by multiplying by the integrating factor (I.F.) = $\exp \left( \int {\rm d}r \left(f'(r) - \lambda\right)/f(r) \right) = (K/2) r^{1-2\alpha} \left( r^{2} - \Delta\right)^{1+\alpha}$, and then integrating with respect to $r$.  Here, we have defined 
\begin{align}			
\alpha \equiv \frac{\lambda
}{2\sigma-K},~~\Delta \equiv 1- \frac{2\sigma}{K}.
\end{align}		
One obtains from Eq.~(\ref{eq:ME-3-st}) the solutions as $P_{\rm {st}}(r)= C_{1,2} ~(\text{I.F.})^{-1}$; in particular,
\begin{eqnarray}
	\begin{split}
		&P_{\rm {st}}(r)= C_{1} \frac{2}{K} r^{2\alpha -1} \left( r^{2} - \Delta\right)^{-(1+\alpha)} \equiv P^{<}_{\rm {st}}(r);~~r \in [0,r_{0}],\\
		&P_{\rm {st}}(r)=C_{2} \frac{2}{K} r^{2\alpha -1} \left( r^{2} - \Delta\right)^{-(1+\alpha)} \equiv P^{>}_{\rm {st}}(r);~~r \in [r_{0},1],
		\label{eq:P_st_gen_soln}
	\end{split}
\end{eqnarray}
with the constants $C_{1}, C_{2}$ to be determined from the normalization condition, Eq.~(\ref{eq:P_normalization}), and the discontinuity at $r=r_{0}$, Eq.~(\ref{eq:P_discont}).  The latter, on using Eq.~(\ref{eq:P_st_gen_soln}), yields 
\begin{align}
	C_{2} -C_{1} = - \lambda \left( 1- \frac{\Delta}{r_{0}^{2}}\right)^{\alpha}.
	\label{eq:coeff_diff}
\end{align}
We note in passing that for a fixed resetting rate $\lambda$, the parameters $\alpha$ and $\Delta$ change sign as one tunes $K$ across $K_{c} = 2\sigma$.  This requires one to exercise caution while normalizing the stationary probability for values of $K$ smaller and larger than $K_c$, which for these two cases are taken up separately in the following.\\

\noindent{\bf Analysis for $K < K_c$}
%\label{sec:KltKc}
\\

In the region $K < K_{c}=2\sigma$, we have $\alpha > 0$ and $\Delta <0 $.  Here,  in  absence of resetting, the dynamics at long times relaxes to a stationary state characterized by $r_{\rm {st}} =0$.  The normalization condition (\ref{eq:P_normalization}) reads
\begin{align}
	\int_{0}^{r_0} {\rm d}r~P^{<}_{\rm {st}}(r) +  \int_{r_0}^{1} {\rm d}r~P^{>}_{\rm {st}}(r)= 1.
	\label{eq:P_norm_unsync}
\end{align}
Using Eq.~(\ref{eq:P_st_gen_soln}),  one obtains
\begin{align}
	&\int_{0}^{r_0} {\rm d}r~P^{<}_{\rm {st}}(r)= \frac{C_{1}}{\lambda} \left( 1- \frac{\Delta}{r_{0}^{2}}\right)^{-\alpha}, \nonumber\\
	\label{eq:norm_int_unsync}\\
	&\int_{r_0}^{1} {\rm d}r~P^{>}_{\rm {st}}(r)=  \frac{C_{2}}{\lambda} \left[ \left( 1- \Delta\right)^{-\alpha} - \left( 1- \frac{\Delta}{r_{0}^{2}}\right)^{-\alpha} \right],\nonumber
\end{align}
which together with Eqs.~(\ref{eq:P_norm_unsync}) and~(\ref{eq:coeff_diff}) yield
\begin{equation}
	C_{1} = \lambda \left( 1- \frac{\Delta}{r_{0}^{2}}\right)^{\alpha}, ~~ C_{2} = 0.
	\label{eq:C1-C2-K<K_c}
\end{equation}

The fact that the coefficient $C_2$ is zero may be understood thus. For $K < K_{c}$ and in the absence of resetting ($\lambda=0$),  the deterministic Kuramoto dynamics~(\ref{eq:eom1}) is such that $r(t)$ while evolving in time from any arbitrary initial condition (including from the one characterized by the given value $r_0$) is attracted towards the stable fixed point $r_{\rm st}=0$ and will thus have for all subsequent times values of $r$ satisfying $r(t>0) \in [0,r_0)$.  However, resetting interrupts the dynamics at random times and brings it back to a state characterized by $r=r_0$.  Then, starting from the initial condition $r=r_0$,  the trajectories $r(t)$ in time that contribute to the probability $P(r,t)$ for a given very large value of $t$ are those that have undergone the last reset at various random time intervals $\tau$ earlier, together with those that have not undergone a single reset since the initial instant $t=0$.   This is because every reset renews the bare Kuramoto dynamics from a state with $r=r_0$.  Among these trajectories,  considering then those that have undergone the last reset just now will have a value of $r$ equal to $r_0$,  those that have undergone the last reset just a while ago will have a value of $r$ slightly less than $r_0$, and continuing this way, those that have undergone the last reset a long time ago will have a value of $r$ close to but larger than $r_{\rm st}=0$.  We thus expect in the limit $t\to \infty$ that the stationary-$r$ will have values in the range $[r_{\rm st}=0, r_{0}]$. The fact that stationary-$r$ does not have a value larger than $r_0$ results in $C_2=0$, as implied by our explicit calculation above.  

The stationary distribution obtained by substituting the expressions of $C_1$ and $C_{2}$ from Eq.~(\ref{eq:C1-C2-K<K_c}) in Eq.~(\ref{eq:P_st_gen_soln}) and omitting in the latter the superscript in the notation for the probability density, we obtain 
\begin{align}
	P_{\rm {st}}(r) =  \frac{2\lambda}{Kr^3} \left( 1+ \frac{|\Delta|}{r_{0}^{2}}\right)^{\alpha} \left( 1 + \frac{|\Delta|}{r^{2}}\right)^{-(1+\alpha)};~~r \in [0,r_0].
	\label{eq:P_st_unsync}
\end{align}

We now investigate the behavior of $P_{\rm {st}}(r)$ near $r_0$, specifically,  as $r \to r_{0}^-$. Taking $r=r_{0} - \epsilon$,   with $0<\epsilon \ll 1$,  one obtains from Eq.~(\ref{eq:P_st_unsync}) in the limit $\epsilon \to 0$ that\begin{align}
	P_{\rm {st}}(r) \sim \frac{2\lambda}{Kr_{0}^{3}} \left(1+ \frac{|\Delta|}{r_{0}^{2}} \right)^{-1},
	\label{eq:P_st_unsync_near_r0}
\end{align}
which shows no dependence on $\epsilon$ to leading order.  Hence,  the stationary distribution has a finite value as $r \to r_{0}^-$ regardless of the value of $\lambda$. This is also consistent with the finite discontinuity obtained in Eq.~(\ref{eq:P_discont}).

\begin{figure}[]
	\centering
	\includegraphics[scale=0.375]{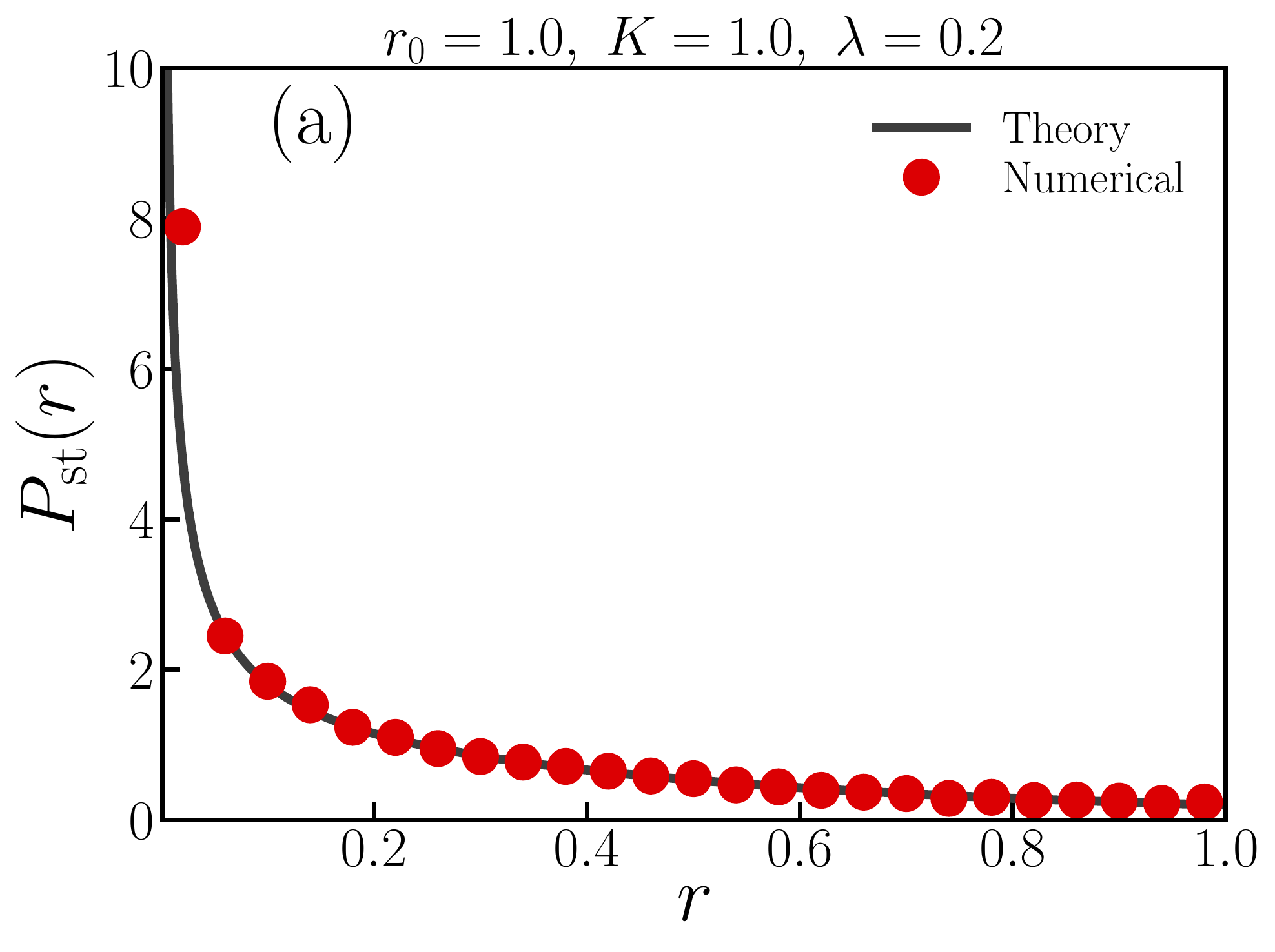}
	\includegraphics[scale=0.38]{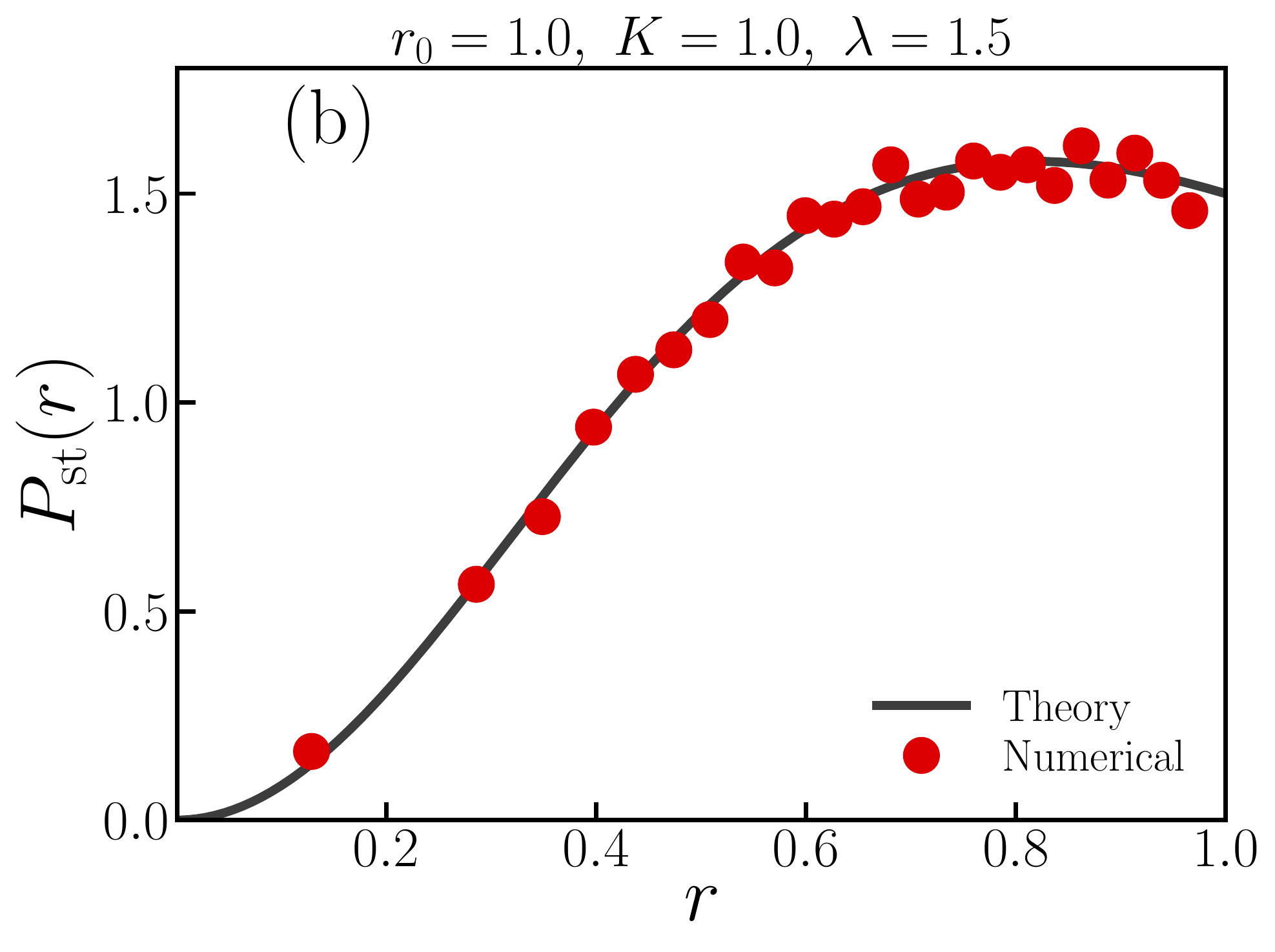}
	\caption{For the Lorentzian distribution of the natural frequencies of the oscillators,  Eq.~(\ref{eq:Gaussian_lorentzian_g_omega}), with width $\sigma=1.0$,  and for the coupling strength $K=1.0<K_c=2\sigma=2.0$,  shown is the stationary-state probability distribution $P_{\rm {st}}(r)$ for values of $\lambda$ smaller and larger than the threshold value $\lambda_c$ given by Eq.~(\ref{eq:lambdac}).  With $\lambda_c=0.5$ in our case,  the chosen values of $\lambda$ are $\lambda = 0.2 < \lambda_c$ [panel (a)] and $\lambda = 1.5 > \lambda_c$ [panel (b)]. The red filled circles, based on $N$-body numerical simulation of the Kuramoto dynamics with resetting for number of oscillators $N=10^4$ and reset parameter $r_0=1.0$,  correspond to the stationary state of the governing dynamics.  The black continuous lines correspond to our theoretical result, Eq.~(\ref{eq:P_st_unsync}), based on the OA ansatz.}
	\label{fig:K_less_Kc_lamb_less_lamb_c}
\end{figure}

Let us now investigate the behavior of $P_{\rm {st}}(r)$ near $r_{\rm st}=0$,  specifically,  as $r \to 0^+$. Take $r=0+ \epsilon$.  In the limit $\epsilon \to 0$, we have from Eq.~(\ref{eq:P_st_unsync}) that
\begin{align}
	P_{\rm {st}}(r) \sim r_{0}^{-2\alpha} \epsilon^{2\alpha -1} \sim \left( \frac{\epsilon}{r_{0}}\right)^{2\alpha -1}.
	\label{eq:P_st_unsync_near_r_0}
\end{align}
At this point, attention may be drawn to the following observations:
Firstly, $P_{\rm {st}}(r)$ diverges as $r \to 0^+$ if we have $2\alpha -1 <0$, i.e., for $\lambda$ smaller than a threshold value 
\begin{align}
	\lambda_{c} \equiv  \sigma \left( 1- \frac{K}{K_{c}} \right).
	\label{eq:lambdac}
\end{align}
Since the probability $P_{\rm st}(r)$ is properly normalized, the aforementioned divergence implies that the distribution becomes very sharply peaked as $r \to 0^+$.  Physically, this means that if the resetting rate is small and in fact less than $\lambda_c$, the resetting dynamics does not effectively compete with the bare Kuramoto dynamics, and hence, the system is most likely to be found in an unsynchronized state as would have been the case if there were no resetting. We find that have $P_{\rm st}(r)$ a monotonically decreasing function in $[0,r_0]$ for $\lambda<\lambda_c$.

Secondly,  we have $P_{\rm {st}}(r) \to 0$ as $r \to 0^+$ for $\lambda > \lambda_c$. In this case,  the competing effects of resetting and bare Kuramoto dynamics result in a peak at $r=r_{\rm m} \equiv \sqrt{(2\alpha -1)|\Delta|/3} >0 $, i.e.,  the most probable value of $r$ is $r_{\rm m}>0$,   provided we have $r_{\rm m}<r_0$.  Physically,  the fact that $P_{\rm {st}}(r) \to 0$ as $r \to 0^+$ for $\lambda > \lambda_c$ implies that one has a vanishing stationary probability of finding the system in the unsynchronized phase.  We thus have that $P_{\rm st}(r)$ is non-monotonic in $[0,r_0]$ for $r_
{\rm m}<r_0$,  and is otherwise monotonically increasing in $[0,r_0]$.

Thirdly, exactly at $\lambda = \lambda_{c}$, $P_{\rm {st}}(r)$ is $\epsilon$-independent as $r \to 0^+$. In fact, one obtains from Eq.~(\ref{eq:P_st_unsync}) for the stationary distribution that
\begin{equation}
	P_{\rm {st}}(r)= \sqrt{1+ \frac{|\Delta|}{r_{0}^{2}}}  \frac{|\Delta|}{\left( r^2 + |\Delta|\right)^{3/2}};~~\lambda = \lambda_{c},
	\label{eq:P_st_unsync_at_lambda_c}
\end{equation}
which approaches a non-zero finite constant as $r \to 0^+$, given by
\begin{equation}
	P_{\rm {st}}(r) \sim \sqrt{\frac{1}{|\Delta|} + \frac{1}{r_{0}^2}};~~\lambda = \lambda_{c},  r \to 0^+.
\end{equation}
We thus have that $P_{\rm st}(r)$ is monotonically decreasing in $[0,r_0]$. 

Our main observations are thus (i) the stationary phase is unsynchronized or synchronized (as determined by the most probable value of stationary-$r$) depending on whether we have $\lambda < \lambda_c$ or $\lambda>\lambda_c$,  respectively,  (ii) the stationary distribution  $P_{\rm {st}}(r)$ is non-zero for $r$ in the range $[r_{\rm st}=0,r_0]$,  (iii) $P_{\rm st}(r)$ as $r \to r_0^-$ is finite for all $\lambda$,  (iv) $P_{\rm {st}}(r)$  for $\lambda < \lambda_c$ is a monotonically decreasing function in $[0,r_{0}]$ with a divergence as $r \to 0^+$,  (v) $P_{\rm {st}}(r)$ for $\lambda > \lambda_c$ is non-monotonic in $[0,r_{0}]$, with the peak at $r_{\rm m}>0 $ provided $r_{\rm m}<r_0$,  while it is monotonically increasing in $[0,r_0]$ for $r_{\rm m}>r_0$,  (vi) $P_{\rm st}(r)$ for $\lambda=\lambda_c$ shows a monotonically decreasing behavior in $[0,r_0]$,  and (vii) $P_{\rm {st}}(r)$ as $r \to 0^+$ is infinite as $\lambda < \lambda_c$, is zero as $\lambda > \lambda_c$,  and is finite at exactly $\lambda=\lambda_c$.  Indeed,  one can write for the behavior of $P_{\rm st}(r)$ at $r=0+\epsilon$ with $0 < \epsilon \ll 1$ as
\begin{align}
P_{\rm st}(r) \sim \epsilon^\beta;~\epsilon \to 0,
\end{align}
where the exponent $\beta$ varies continuously with the deviation $\lambda-\lambda_c$:
\begin{align}
\beta=\beta(\lambda-\lambda_c).
\end{align}
Here, the function $\beta(x)$ has the following behavior:
\begin{align}
\beta(0)=0, ~\beta(x<0)<0, ~\beta(x>0)>0.
\end{align}
We remark that a distinguishing feature of the threshold $\lambda_c$ given by Eq.~(\ref{eq:lambdac}) is that as one tunes $\lambda$ at a fixed $K$ from smaller to larger values across $\lambda_c$, the stationary distribution $P_{\rm st}(r)$ shows a divergence as $r \to r_{\rm st}^+$ for $\lambda < \lambda_c$,  has a finite value at $\lambda=\lambda_c$, and is zero for $\lambda>\lambda_c$.

Based on the above discussion,  we conclude that for fixed $K<K_c$,  when the bare Kuramoto dynamics does not allow for any synchronized phase in the stationary state,   introducing resetting leads to the dramatic emergence of a ``pseudo-synchronized phase'': on tuning the resetting rate $\lambda$ from low to high values across the threshold value~(\ref{eq:lambdac}), one goes over from an unsynchronized phase to the pseudo-synchronized phase.  The emergence of the latter phase is intimately tied to the introduction of resetting in the dynamics. Indeed, this phase vanishes for $\lambda=0$ (no resetting), as follows from the fact that the threshold line given by Eq.~(\ref{eq:lambdac}) in the $(K,\lambda)$-plane meets the $K$ axis at $K=K_c$, see Fig.~\ref{fig:phase-diagram}.

The theoretical result~(\ref{eq:P_st_unsync}), which is based on the OA ansatz, is checked in Fig.~\ref{fig:K_less_Kc_lamb_less_lamb_c} against results based on numerical simulation of the Kuramoto dynamics with resetting performed with number of oscillators $N=10^4$ and a Lorentzian distribution of the natural frequencies, Eq.~(\ref{eq:Gaussian_lorentzian_g_omega}), with width $\sigma=1.0$.  We choose the coupling constant to be $K=1.0$,  while the reset parameter is $r_0=1.0$. We consider the resetting rate $\lambda$ to be both smaller and larger than the threshold value~(\ref{eq:lambdac}); as discussed above and as may be seen from the figure, the stationary distribution $P_{\rm st}(r)$ does exhibit distinctly different behavior depending on whether $\lambda$ is smaller or larger than $\lambda_c$.\\

\noindent{\bf Analysis for $K > K_c$} 
%\label{sec:KgtKc}
\\

For $K > K_{c}$, we have $\alpha < 0$ and $\Delta >0$.  Here,  in  absence of resetting, the dynamics at long times relaxes to a stationary state characterized by $r_{\rm {st}} = \sqrt{1-K_c /K}$, with $K_c=2\sigma$ and $ 0 < r_{\rm st} \le 1$.  Note that in this region, we have $\Delta = r_{\rm{st}}^{2}$.  Following the same line of argument as adduced above for $K<K_c$,  one can conclude that in presence of resetting, the stationary-$r$ will lie in either the range $[r_0, r_{\rm{st}}]$, or, the range $[r_{\rm{st}}, r_0]$, depending on whether $r_0 < r_{\rm{st}}$, or, $r_0 > r_{\rm{st}}$, respectively. Then,  for $r_0>r_{\rm{st}}$, we have $C_{2} =0$, and the normalization condition~(\ref{eq:P_normalization}) reads
\begin{align}
	\int_{r_{\rm{st}}}^{r_0} {\rm d}r~P^{<}_{\rm {st}}(r) = 1,
\end{align}
which together with Eq.~(\ref{eq:P_st_gen_soln}) yields 
\begin{align}
	P_{\rm {st}}(r) =  \frac{2\lambda}{Kr^3} \left( 1- \frac{r_{\rm{st}}^{2}}{r_{0}^{2}}\right)^{-|\alpha|} \left(1 - \frac{r_{\rm{st}}^{2}}{r^{2}}\right)^{|\alpha|-1};~~r \in [r_{\rm{st}}, r_{0}].
	\label{eq:P_st_sync_I}
\end{align}
On the other hand,  for $r_0 < r_{\rm{st}}$,  one has $C_{1} =0$, and one obtains on using 
\begin{align}
	\int^{r_{\rm{st}}}_{r_0} {\rm d}r~P^{>}_{\rm {st}}(r) = 1
\end{align}
that
\begin{align}
	P_{\rm {st}}(r) =  \frac{2\lambda}{Kr^3} \left(  \frac{r_{\rm{st}}^{2}}{r_{0}^{2}} -1\right)^{-|\alpha|} \left( \frac{r_{\rm{st}}^{2}}{r^{2}} -1\right)^{|\alpha|-1};~~r \in [r_{0},r_{\rm{st}}].
	\label{eq:P_st_sync_II}
\end{align}
%{\color{blue}Since the stationary probability $P_{\rm {st}}(r)$ is non-zero only over a non-zero range of values of $r$, it immediately follows that  there will always be a synchronized phase for $K> K_c$ regardless of the value of the resetting rate $\lambda$. }

Considering the case $r_0<r_{\rm{st}}$,  let us now study the behavior of $P_{\rm {st}}(r)$ as $r\to r_0^+$. Taking $r=r_0 + \epsilon$,  with $ 0 < \epsilon \ll 1$,  one obtains from Eq.~(\ref{eq:P_st_sync_II}) in the limit $\epsilon \to 0$ that\begin{align}
	P_{\rm {st}}(r) \sim \frac{2\lambda}{Kr_{0}^{3}} \left( \frac{r_{\rm{st}}^{2}}{r_{0}^{2}} -1\right)^{-1},
	\label{eq:P_st_sync_near_r_0}
\end{align}
which shows no dependence on $\epsilon$ to leading order.  Hence,  the stationary distribution has a finite value as $r \to r_{0}^+$ regardless of the value of $\lambda$. This is also consistent with the finite discontinuity given by Eq.~(\ref{eq:P_discont}).  

We now study the behavior of $P_{\rm {st}}(r)$ near $r=r_{\rm{st}}$. 
Take $r=r_{\rm{st}} -\epsilon$. In the limit $\epsilon \to 0$, i.e., $r \to r^{-}_{\rm{st}}$, one obtains from Eq.~(\ref{eq:P_st_sync_II}) that
\begin{align}
	P_{\rm {st}}(r) \sim \frac{1}{r_{\rm{st}}^3} \left( \frac{\epsilon}{r_{\rm{st}}} \right)^{|\alpha|-1}.
	\label{eq:P_st_sync_near_r_st}
\end{align}
One thus observes that if we have $|\alpha|-1 <0$, i.e., $\lambda < K-K_{c}$, the distribution $P_{\rm {st}}(r)$ peaks sharply at $r \to r^{-}_{\rm{st}}$, implying thereby that in this regime of $\lambda$,  the effects of resetting are subdominant with respect to those of inter-oscillator interactions. In such a case, depending on the reset value $r_0$, the distribution may have a dip at $r=r_{\rm d}\equiv \sqrt{\left(2 |\alpha| + 1\right)/3}~ r_{\rm {st}}$: if $r_0 < r_{\rm d}$, $P_{\rm {st}}(r)$ shows the dip at $r=r_{\rm d} < r_{\rm {st}}$, showing a non-monotonic behavior in $[r_0, r_{\rm {st}}]$, otherwise it is a monotonically increasing function in $[r_0, r_{\rm {st}}]$. 

\begin{figure}[]
	\centering
	\includegraphics[scale=0.375]{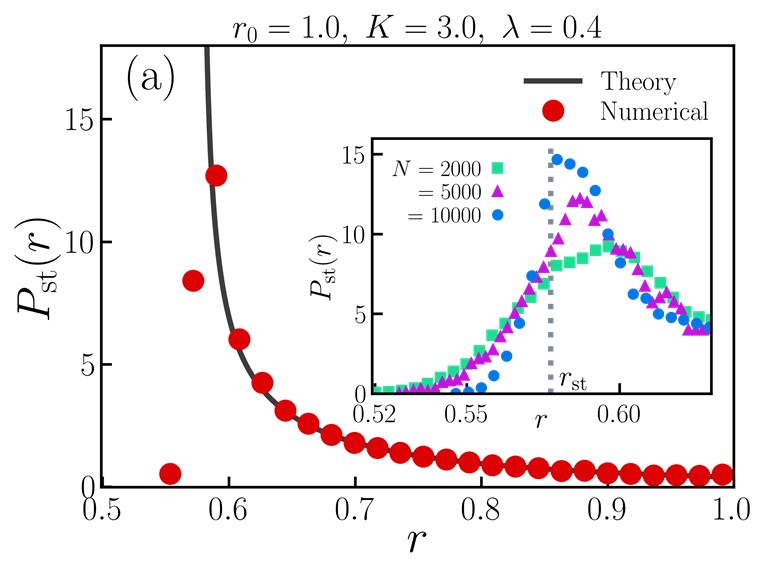}
	\includegraphics[scale=0.375]{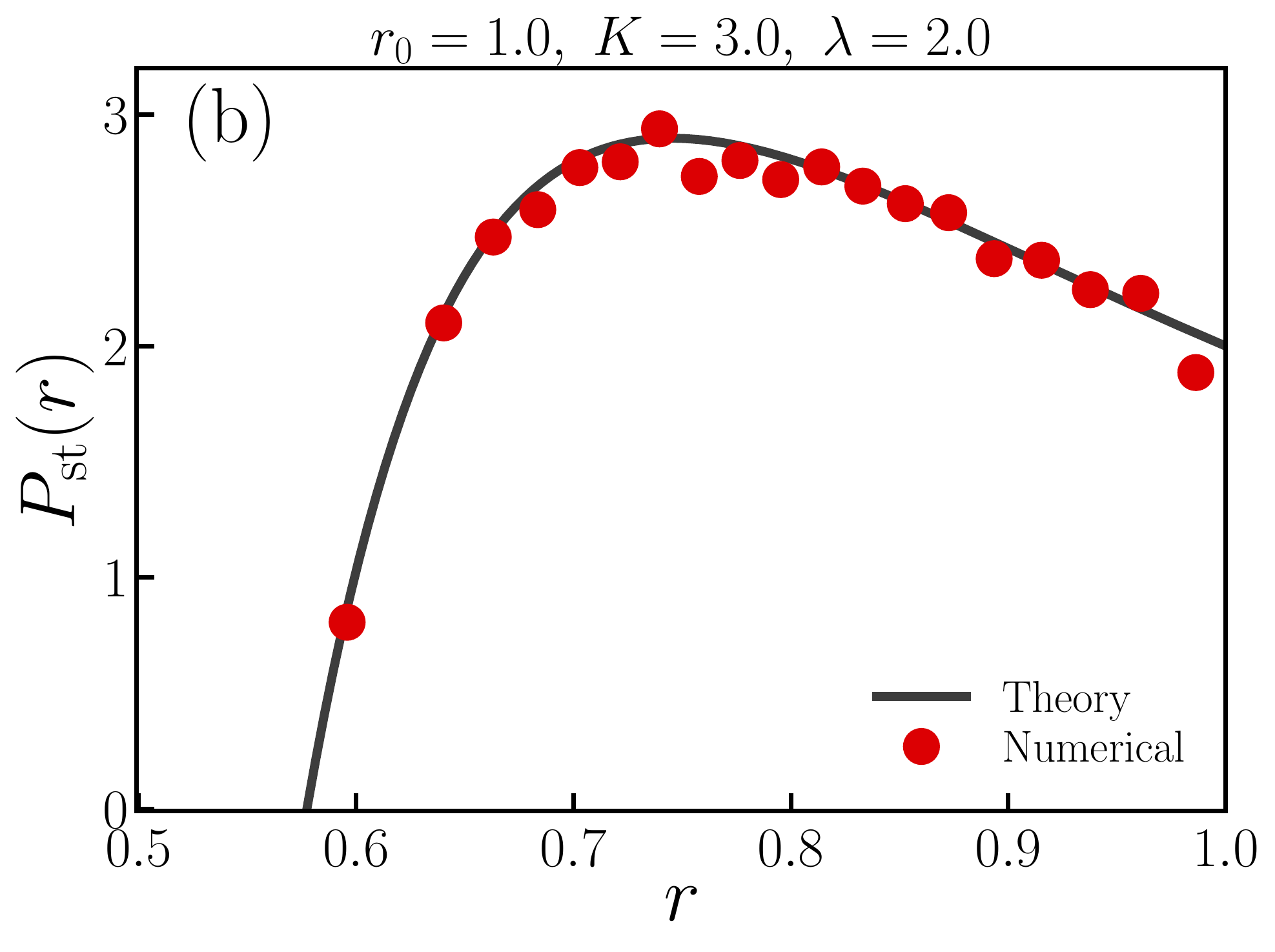}
	\caption{For the Lorentzian distribution of the natural frequencies of the oscillators,  Eq.~(\ref{eq:Gaussian_lorentzian_g_omega}), with width $\sigma=1.0$,  and for the coupling strength $K=3.0>K_c=2\sigma=2.0$,  shown is the stationary-state probability distribution $P_{\rm {st}}(r)$ for values of $\lambda$ smaller and larger than $\lambda^\star=K-K_c$.  With $\lambda^\star=1.0$ in our case,  the chosen values of $\lambda$ are $\lambda = 0.4 < \lambda^\star$ [panel (a)] and $\lambda = 2.0 > \lambda^\star$ [panel (b)]. The red filled circles, based on $N$-body numerical simulation of the Kuramoto dynamics with resetting for number of oscillators $N=10^4$ and reset parameter $r_0=1.0$,  correspond to the stationary state of the governing dynamics.  Note that the choice of $K$ and $r_0$ implies that we have $r_{0} > r_{\rm {st}}$. The black continuous lines correspond to our theoretical result, Eq.~(\ref{eq:P_st_sync_I}), based on the OA ansatz.  In panel (a), our theory predicts a divergence of the distribution function $P_{\rm {st}}(r)$ as $r \to r^+_{\rm {st}}$, while in the same region,  we observe in simulations a finite-size rounding off resulting in a finite $P_{\rm {st}}(r)$. Moreover, unlike in theory,  we observe in simulations non-zero values of $P_{\rm {st}}(r)$ for $r < r_{\rm {st}}$.  Increasing conformity to theoretical results ($P_{\rm st}(r)$ diverging as $r \to r^+_{\rm {st}}$ and is zero for $r < r_{\rm {st}}$) with an increase in the number of oscillators $N$ may be seen in the inset of panel (a).}
	\label{fig:K_less_Kc_lamb_gt_lamb_c}
\end{figure}

\begin{figure}[]
	\centering
	\includegraphics[scale=0.375]{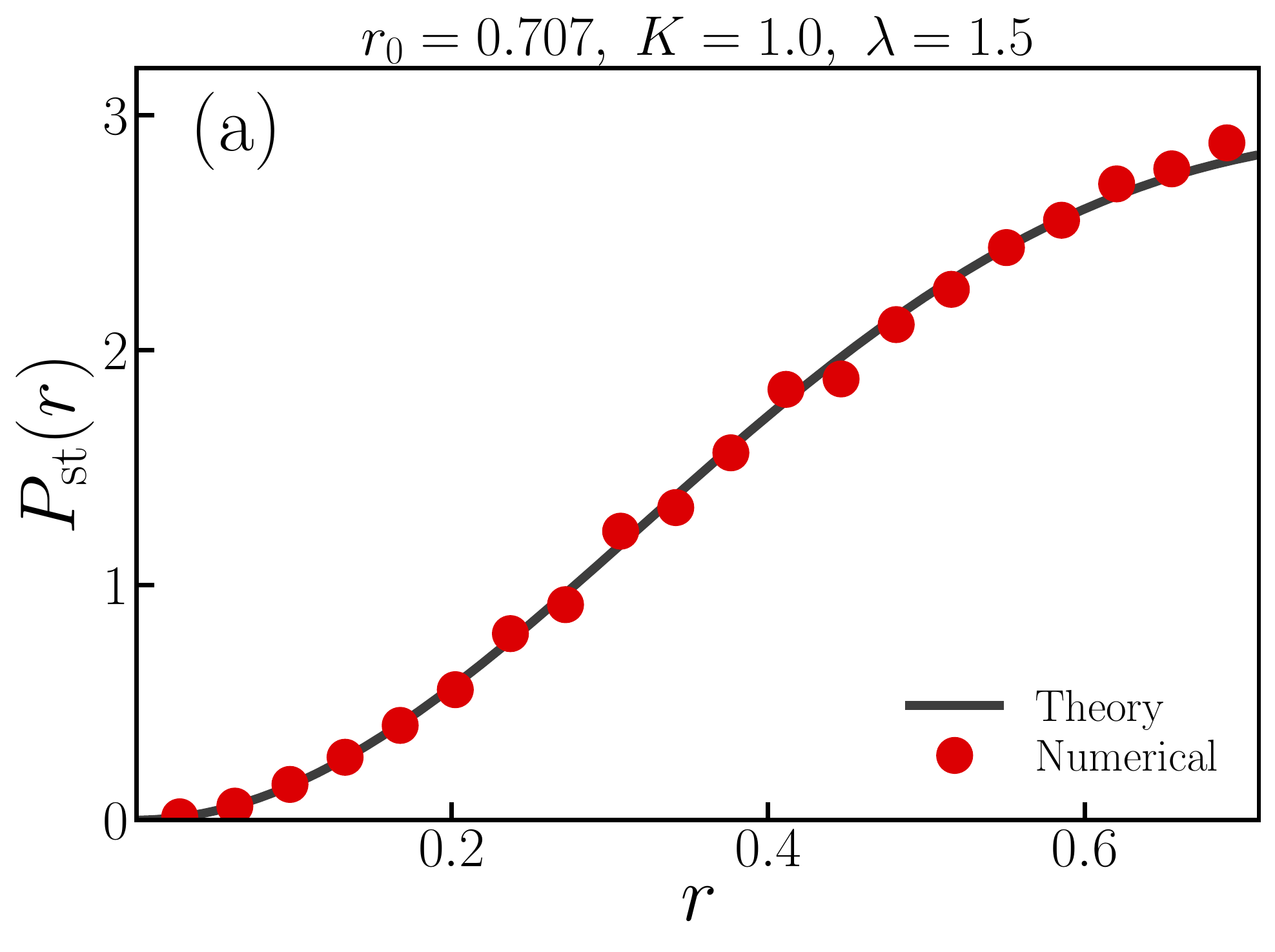}
	\includegraphics[scale=0.375]{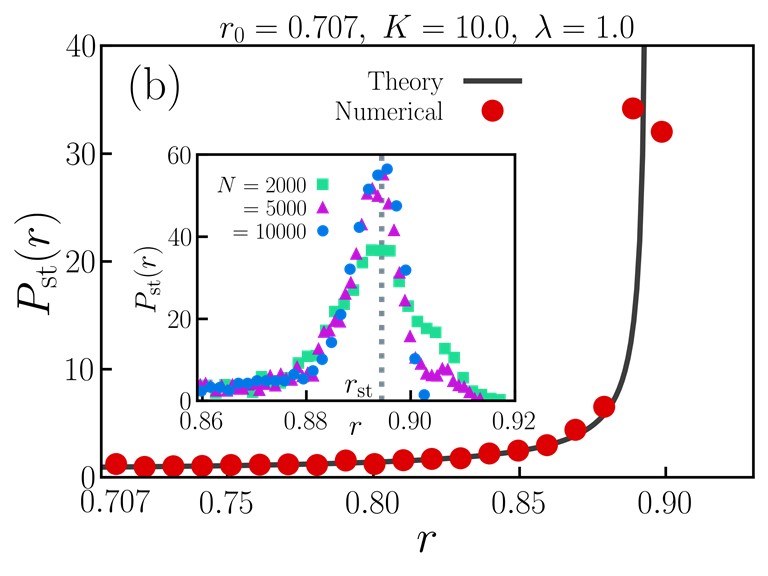}
	\caption{For the Lorentzian distribution of the natural frequencies of the oscillators, Eq.~(\ref{eq:Gaussian_lorentzian_g_omega}), with width $\sigma=1.0$, shown is the stationary-state probability distribution $P_{\rm {st}}(r)$ for the coupling strength $K=1.0<K_c=2\sigma=2.0$ and $\lambda=1.5$ [panel (a)], and for $K=10.0>K_c=2\sigma=2.0$ and $\lambda=1.0$ [panel (b)], with the reset parameter $r_0=0.707$. The red filled circles, based on $N$-body numerical simulation of the Kuramoto dynamics with resetting for number of oscillators $N=10^4$ correspond to the stationary state of the governing dynamics. The black continuous lines correspond to our theoretical result based on the OA ansatz: Eq.~(\ref{eq:P_st_unsync}) for panel (a) and Eq.~(\ref{eq:P_st_sync_II}) for panel (b). In panel (b), our theory predicts a divergence of the distribution function $P_{\rm {st}}(r)$ as $r \to r^-_{\rm {st}}$, while in the same region,  we observe in simulations a finite-size rounding off resulting in a finite $P_{\rm {st}}(r)$. Moreover, unlike in theory,  we observe in simulations non-zero values of $P_{\rm {st}}(r)$ for $r > r_{\rm {st}}$.  Increasing conformity to theoretical results ($P_{\rm st}(r)$ diverging as $r \to r^-_{\rm {st}}$ and is zero for $r > r_{\rm {st}}$) with an increase in the number of oscillators $N$ may be seen in the inset of panel (b).}
	\label{fig:P_st_r0_0p7}
\end{figure}

\begin{figure}[]
	\centering
	\includegraphics[scale=0.375]{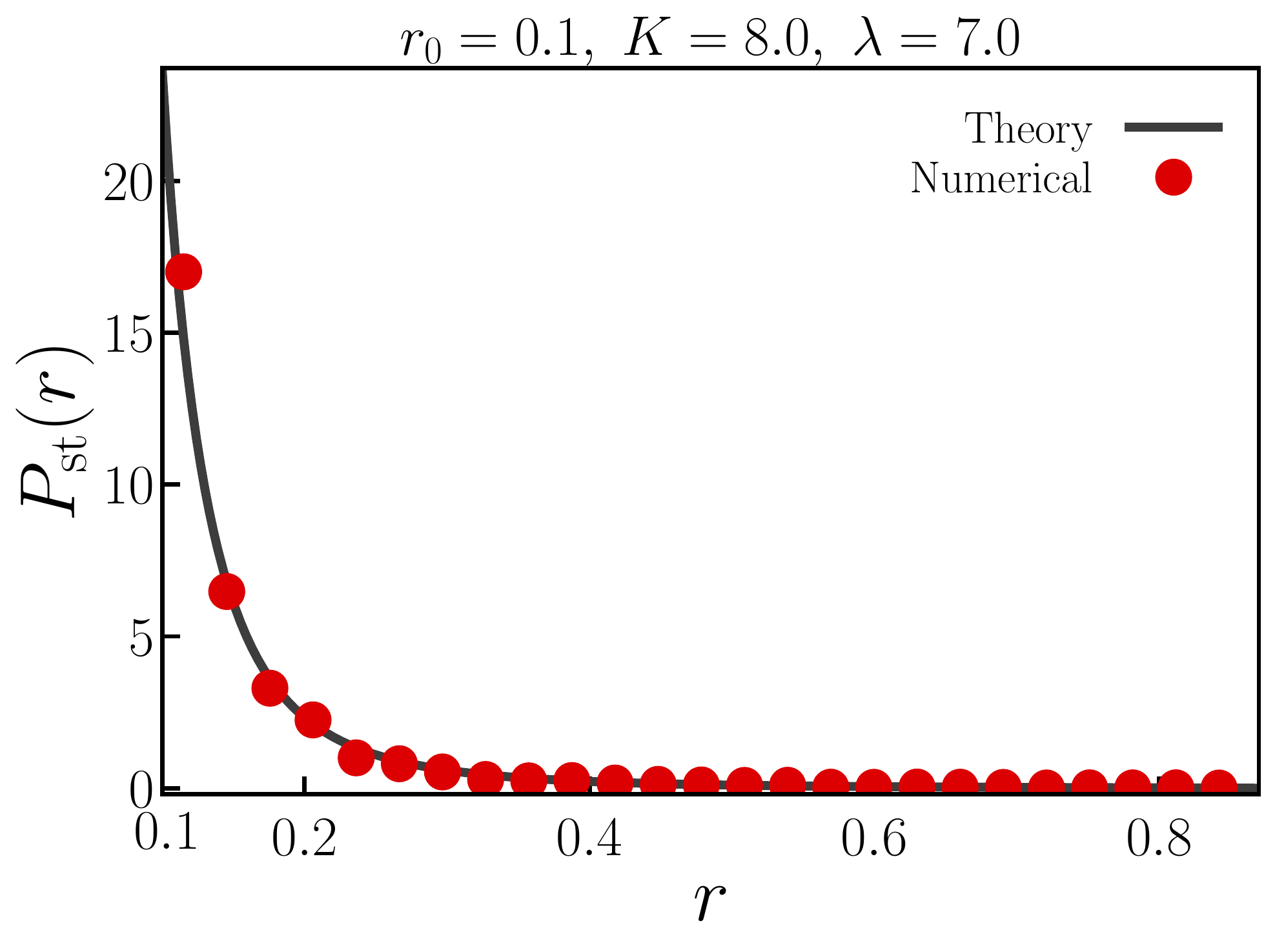}
	\caption{For the Lorentzian distribution of the natural frequencies of the oscillators, Eq.~(\ref{eq:Gaussian_lorentzian_g_omega}), with width $\sigma=1.0$, shown is the stationary-state probability distribution $P_{\rm {st}}(r)$ for the coupling strength $K=8.0>K_c=2\sigma=2.0$ and $\lambda=7.0$.  The reset parameter has the value $r_0=0.1$. The red filled circles, based on $N$-body numerical simulation of the Kuramoto dynamics with resetting for number of oscillators $N=10^4$ correspond to the stationary state of the governing dynamics. The black continuous lines correspond to our theoretical result based on the OA ansatz and given by Eq.~(\ref{eq:P_st_sync_II}).}
	\label{fig:P_st_r0_0p1}
\end{figure}

Now,  we come to the behavior for $\lambda > K-K_c$. Here,  we have $P_{\rm {st}}(r) \to 0$ as $r \to r^{-}_{\rm{st}}$, implying that now, resetting effects dominate, and we find that $P_{\rm {st}}(r)$ is a monotonically decreasing function in $[r_0, r_{\rm {st}}]$. Note that now,  $r_{\rm d}$ has a value outside the admissible range $[r_0,r_{\rm st}]$. 

Exactly at $\lambda=K-K_c$, one has $r_{\rm d}=r_{\rm st}$ and that as $r \to r^{-}_{\rm{st}}$ the behavior $P_{\rm {st}}(r) \to$ a non-zero finite value that is $\epsilon$-independent. Indeed, one obtains from Eq.~(\ref{eq:P_st_sync_II}) that
\begin{equation}
		P_{\rm {st}}(r) \sim  \frac{2\lambda}{Kr^3} \left(  \frac{r_{\rm{st}}^{2}}{r_{0}^{2}} -1\right)^{-1};~~\lambda = K-K_c,
\end{equation}
which approaches $\sim {2\lambda}/({Kr_{\rm{st}}^3}) \left(  {r_{\rm{st}}^{2}}/{r_{0}^{2}} -1\right)^{-1}$
as $r \to r^{-}_{\rm{st}}$. In this case, the distribution $P_{\rm {st}}(r)$ is a monotonically decreasing function in $[r_0, r_{\rm {st}}]$.

A similar analysis as above may be carried out for $r_0>r_{\rm st}$ and the behavior of $P_{\rm {st}}(r)$ near $r = r_0$ and $r=r_{\rm{st}}$ may be investigated. Here, the stationary distribution has a finite value $\sim 2\lambda/(Kr_{0}^{3})\,(1- r_{\rm{st}}^{2}/r_{0}^{2})^{-1}$ as $r \to r_{0}^-$ regardless of the value of $\lambda$.  Next, as $r \to r_{\rm{st}}^+$, we observe that Eq.~(\ref{eq:P_st_sync_near_r_st}) holds in this case too.  If $\lambda < K-K_c$, the distribution $P_{\rm {st}}(r)$ peaks sharply at $r \to r^{+}_{\rm{st}}$, whereas it tends to zero as $r \to r^{+}_{\rm{st}}$ for $\lambda > K-K_c$. However, in contrast to the previously-studied case ($r_0<r_{\rm{st}}$), in the region $\lambda < K-K_c$, we have $P_{\rm {st}}(r)$ a monotonically decreasing function in $[r_{\rm {st}}, r_0]$, while for $\lambda > K-K_c$, there may be a peak in the stationary distribution of $r$ at $r=r_{\rm p} \equiv \sqrt{\left(2 |\alpha| + 1\right)/3}~ r_{\rm {st}}$ depending on $r_0$. If $r_0 > r_{\rm p}$, $P_{\rm {st}}(r)$ has a peak at a value $r_{\rm p}> r_{\rm{st}}$, showing a non-monotonic behavior in $[r_{\rm {st}}, r_0]$, otherwise it shows a monotonically increasing behavior in $[r_{\rm {st}}, r_0]$.  Exactly at $\lambda=K-K_c$,  one has $r_{\rm p}=r_{\rm st}$ and that as $r \to r^{+}_{\rm{st}}$ the behavior $P_{\rm {st}}(r) \to$ a non-zero finite value that is $\epsilon$-independent, given by $P_{\rm {st}}(r) \sim {2\lambda}/({Kr_{\rm{st}}^3}) \left( 1- {r_{\rm{st}}^{2}}/{r_{0}^{2}} \right)^{-1}$.  In this case, $P_{\rm {st}}(r) = {2\lambda}/({Kr^3}) \left( 1- {r_{\rm{st}}^{2}}/{r_{0}^{2}} \right)^{-1}$ is a monotonically decreasing function in $[r_{\rm st}, r_0]$. 

Our main observations are thus (i) the stationary phase is a synchronized one (as determined by the most probable value of stationary-$r$) regardless of the value of $\lambda$, (ii) the stationary distribution  $P_{\rm {st}}(r)$ is non-zero for $r$ in the range $[r_0,r_{\rm st}]$, or,  in the range $[r_{\rm st},r_0]$, depending on whether we have (a) $r_0<r_{\rm st}$, or, (b) $r_0>r_{\rm st}$,  respectively, 
(iii) $P_{\rm st}(r)$ as $r \to r_0^+$ for (a) and as $r \to r_0^-$ for (b) is finite for all $\lambda$,  (iv) $P_{\rm {st}}(r)$ for $\lambda < \lambda^\star$, with
\begin{align}
\lambda^\star 
 \equiv (K - K_c),
 \label{eq:lambdastar} 
 \end{align}
 is for (a) non-monotonic with a dip at $r_{\rm d}$ for $r_0 < r_{\rm d}$, or, a monotonically increasing function of $r$ for $r_0 > r_{\rm d}$, and for (b) always a monotonically decreasing function of $r$,
(v) $P_{\rm {st}}(r)$ for $\lambda > \lambda^\star$ is always a monotonically decreasing function of $r$ for (a), while for (b) it is non-monotonic with a peak at a value $r_{\rm p}$ for the case $r_0 > r_{\rm p}$ and monotonically increasing for $r_0 < r_{\rm p}$, (vi) $P_{\rm {st}}(r)$ for $\lambda = \lambda^\star$ shows a monotonically decreasing behavior for both (a) and (b) in their respective allowed range of values of $r$, and (vii) $P_{\rm {st}}(r)$ as $r \to r_{\rm st}^-$ for (a) and as $r \to r_{\rm st}^+$ for (b) diverges for $\lambda < \lambda^\star$, is zero for $\lambda > \lambda^\star$,  and is finite at exactly $\lambda=\lambda^\star$.  Indeed,  one can write for the behavior of $P_{\rm st}(r)$ at $r=r_{\rm st}-\epsilon$ for (a) and at $r=r_{\rm st}+\epsilon$ for (b), with $0 < \epsilon \ll 1$, as
\begin{align}
P_{\rm st}(r) \sim \epsilon^{\widetilde{\beta}};~\epsilon \to 0,
\end{align}
where the exponent $\widetilde{\beta}$ varies continuously with the deviation $\lambda-\lambda^\star$:
\begin{align}
\widetilde{\beta}=\widetilde{\beta}(\lambda-\lambda^\star);~~\lambda^\star = (K-K_c).
\end{align}
Here, the function $\widetilde{\beta}(x)$ has the following behavior:
\begin{align}
\widetilde{\beta}(0)=0, ~\widetilde{\beta}(x<0)<0, ~\widetilde{\beta}(x>0)>0.
\end{align}
Similar to what we had for $\lambda_c$,  we remark that a distinguishing feature of the threshold $\lambda^\star$ given by Eq.~(\ref{eq:lambdastar}) is that as one tunes $\lambda$ at a fixed $K$ from smaller to larger values across $\lambda^\star$, the stationary distribution $P_{\rm st}(r)$ shows a divergence as $r \to r_{\rm st}^\pm$ for $\lambda < \lambda^\star$, has a finite value at $\lambda=\lambda^\star$, and is zero for $\lambda>\lambda^\star$.    

The theoretical result~(\ref{eq:P_st_sync_I}), which is based on the OA ansatz, is checked in Fig.~\ref{fig:K_less_Kc_lamb_gt_lamb_c} against results based on numerical simulation of the Kuramoto dynamics with resetting performed with number of oscillators $N=10^4$ and a Lorentzian distribution of the natural frequencies, Eq.~(\ref{eq:Gaussian_lorentzian_g_omega}), with width $\sigma=1.0$.  We choose the coupling constant to be $K=3.0$,  while the reset parameter is $r_0=1.0$.   The choice of $K$ and $r_0$ implies that we have $r_0 > r_{\rm st}$.  We consider the resetting rate $\lambda$ to be both smaller and larger than $\lambda^\star=(K-K_c)$.  As discussed above and as may be seen from the figure, the stationary distribution $P_{\rm st}(r)$ does exhibit distinctly different behavior for $\lambda$ smaller and larger than $\lambda^\star$.

While Fig.~\ref{fig:K_less_Kc_lamb_less_lamb_c} and Fig.~\ref{fig:K_less_Kc_lamb_gt_lamb_c} validate our analytical results for $P_{\rm st}(r)$ for the case $r_0=1.0$, we show in Figs.~\ref{fig:P_st_r0_0p7} and~\ref{fig:P_st_r0_0p1}  that the validity holds also for smaller values of $r_0$. In Fig.~\ref{fig:P_st_r0_0p7}, we have $r_0=0.707$, while for Fig.~\ref{fig:P_st_r0_0p1}, we have $r_0=0.1$; the values of other parameters are indicated in the figures.

Let us conclude this part by emphasizing that, unlike for $K<K_c$, here one does not have any new phase emerging due to resetting.   Specifically, one always has a synchronized phase for $K>K_c$, regardless of the value of the resetting rate $\lambda$.\\

\noindent{\bf Analysis at $K=K_c$}\\
%\label{sec:KeqKc}

Here,  to obtain our results,  we need to study separately the cases $K \to K_c^{-}$ and $K \to K_c^{+}$ by using the results obtained above.  To study the former case, take $K=K_c-\tilde{\delta}$ with $0<\tilde{\delta} \ll 1$. In the limit $\tilde{\delta} \to 0$, i.e., as $K \to K_c^{-}$, one can easily show from Eq.~(\ref{eq:P_st_unsync}) for the stationary distribution that 
\begin{equation}
	P_{\rm {st}}(r) = \frac{2\lambda}{K_{c}r^{3}} \exp \left[ -\frac{\lambda}{K_c} \left( \frac{1}{r^2} - \frac{1}{r_0^2}\right)\right]; ~~ r \in [0,r_0].
	\label{eq:P_st_K_Kc_minus}
\end{equation}
Let us now investigate the behavior near $r=r_0$ and $r=0$. Take $r=r_0 - \epsilon$,  with $ 0 < \epsilon \ll 1$. We have from Eq.~(\ref{eq:P_st_K_Kc_minus}) in the limit $\epsilon \to 0$, i.e. , as $r \to r_0^{-}$, that
\begin{equation}
P_{\rm {st}}(r) \sim \frac{2\lambda}{K_{c}r_0^{3}},
\end{equation}
an $\epsilon$-independent non-zero finite constant. On the other hand, taking $r=0 + \epsilon$,  with $ 0 < \epsilon \ll 1$, one obtains from Eq.~(\ref{eq:P_st_K_Kc_minus}) in the limit $\epsilon \to 0$, i.e., as $r \to 0^+$, that 
\begin{equation}
	P_{\rm {st}}(r) \sim \frac{2\lambda}{K_{c}\epsilon^{3}} \exp \left[ -\frac{\lambda}{K_c \epsilon^2} \right],
\end{equation}
which tends to $0$ as $\epsilon \to 0$. In $[0, r_0]$, the distribution $P_{\rm {st}}(r)$
is non-monotonic with a peak at $r^* = \sqrt{2 \lambda/(3K_c)}$ if $r^* < r_0$; otherwise, it is a monotonically increasing function in $[0, r_0]$. 

We will now investigate the behavior of $P_{\rm {st}}(r)$ in the limit $K \to K_c^{+}$. Take $K=K_c + \tilde{\delta}$ with $0<\tilde{\delta} \ll 1$. As we know, in the limit $\tilde{\delta} \to 0$, i.e., as $K \to K_c^{+}$,  and in absence of resetting, the stationary order parameter satisfies $r_{\rm {st}}=\sqrt{1-(2\sigma/K)} \approx \sqrt{\tilde{\delta}/K_c}$,  so we consider only the case $r_{\rm {st}} < r_0$. As $K \to K_c^{+}$, one obtains from Eq.~(\ref{eq:P_st_sync_I}) a behavior similar to Eq.~(\ref{eq:P_st_K_Kc_minus}):
\begin{equation}
	P_{\rm {st}}(r) = \frac{2\lambda}{K_{c}r^{3}} \exp \left[ -\frac{\lambda}{K_c} \left( \frac{1}{r^2} - \frac{1}{r_0^2}\right)\right]; ~~ r \in [r_{\rm {st}},r_0].
	\label{eq:P_st_K_Kc_plus}
\end{equation}
One immediately has from Eq.~(\ref{eq:P_st_K_Kc_plus}) that $P_{\rm {st}}(r) \sim {2\lambda}/{K_{c}r_0^{3}}$ as $r \to r_0^-$. On the other hand, as $r \to r_{\rm {st}}^{+} $, one has \begin{equation}
	P_{\rm {st}}(r) \sim \frac{2\lambda \sqrt{K_{c}}}{{(K-K_c)}^{3/2}} \exp \left[ -\frac{\lambda}{{K-K_c}} \right].
\end{equation} 
We find that $P_{\rm {st}}(r)$ shows a non-monotonic behavior in $[r_{\rm {st}}, r_0]$ with a peak at $r^* = \sqrt{2 \lambda/(3K_c)}$ if $r^* < r_0$; otherwise, it is a monotonically increasing function in $[r_{\rm {st}}, r_0]$.

\subsection{Phase diagram}
\label{sec:phase-diagram}
For the case $r_0=0$, using Eq.~(\ref{eq:ME-st-r0-0}), we conclude that the phase diagram in the $(K,\lambda)$-plane is trivial: the system is always unsynchronized. The first moment, i.e., the mean $r_{\rm {st}}^{(\lambda)}$ of the stationary-$r$ distribution $P_{\rm st}(r)$, defined as $r_{\rm {st}}^{(\lambda)} \equiv \int {\rm d}r ~ r P_{\rm st}(r)$, is zero everywhere in the $(K,\lambda)$-plane.

We now turn to the case $0 < r_0 \leq 1$.
The complete phase diagram in the $(K,\lambda)$-plane is shown in Fig.~\ref{fig:phase-diagram}.  The two axes are the coupling constant $K \ge 0$ and the resetting rate $\lambda \ge 0$.  The whole of the shaded regions corresponds to a synchronized phase, while the white region in the bottom left refers to an unsynchronized phase. This is in the sense that the most probable value of $r_\mathrm{m}$, given by the value of $r$ at which the distribution $P_\mathrm{st}(r)$ peaks, is nonzero in the shaded regions and is zero in the white region. In the diagram, we also show via the different insets the wide spectrum of different behavior exhibited by the stationary-state order parameter distribution $P_{\rm st}(r)$.  The dash-dotted line corresponds to $K=K_c$.  The solid line denotes the threshold $\lambda_c$ given by Eq.~(\ref{eq:lambdac}), while the dashed line denotes the threshold $\lambda^\star$ given by Eq.~(\ref{eq:lambdastar}). Above the lines for $\lambda_c$ and $\lambda^\star$, the dynamics is resetting dominated, which manifests in $P_{\rm st}(r) \to 0$ (i) as $r \to 0^+$ for $K<K_c$, and (ii) as $r \to r_{\rm st}^\pm$ for $K>K_c$. Note that $r_{\rm st}=0$ for $K \le K_c$ and $r_{\rm st}=\sqrt{1-K_c/K}$ for $K \ge K_c$.  Below these lines,  when the bare Kuramoto dynamics~(\ref{eq:eom1}) is dominating, we have $P_{\rm st}(r)$ diverging (i) as $r \to 0^+$ for $K<K_c$, and (ii) as $r \to r_{\rm st}^\pm$ for $K>K_c$.  The stationary distribution $P_{\rm st}(r)$ for values of $\lambda$ above $\lambda_c$ and $\lambda^\star$ is characterized by either a monotonic dependence on $r$ or a non-monotonic dependence, with a maximum at $r_{\rm m}$ (for $\lambda > \lambda_c$) and $r_{\rm p}$ (for $\lambda > \lambda^\star$).  For $\lambda < \lambda_c$,  one has only a monotonic dependence on $r$,  in contrast to the case for $\lambda < \lambda^\star$ when one may have either a monotonic dependence or a non-monotonic one with a minimum at $r_{\rm d}$.  Note that the qualitative features of $P_{\rm st}(r)$ change on crossing either the dash-dotted or the solid or the dashed line.  However,  the solid line is special with respect to the other two in that it marks the boundary between a synchronized and an unsynchronized phase;  on the other hand,  one has a synchronized phase on either side of the other two lines. 

\begin{figure}[!ht]
	\centering
	\includegraphics[scale=0.375]{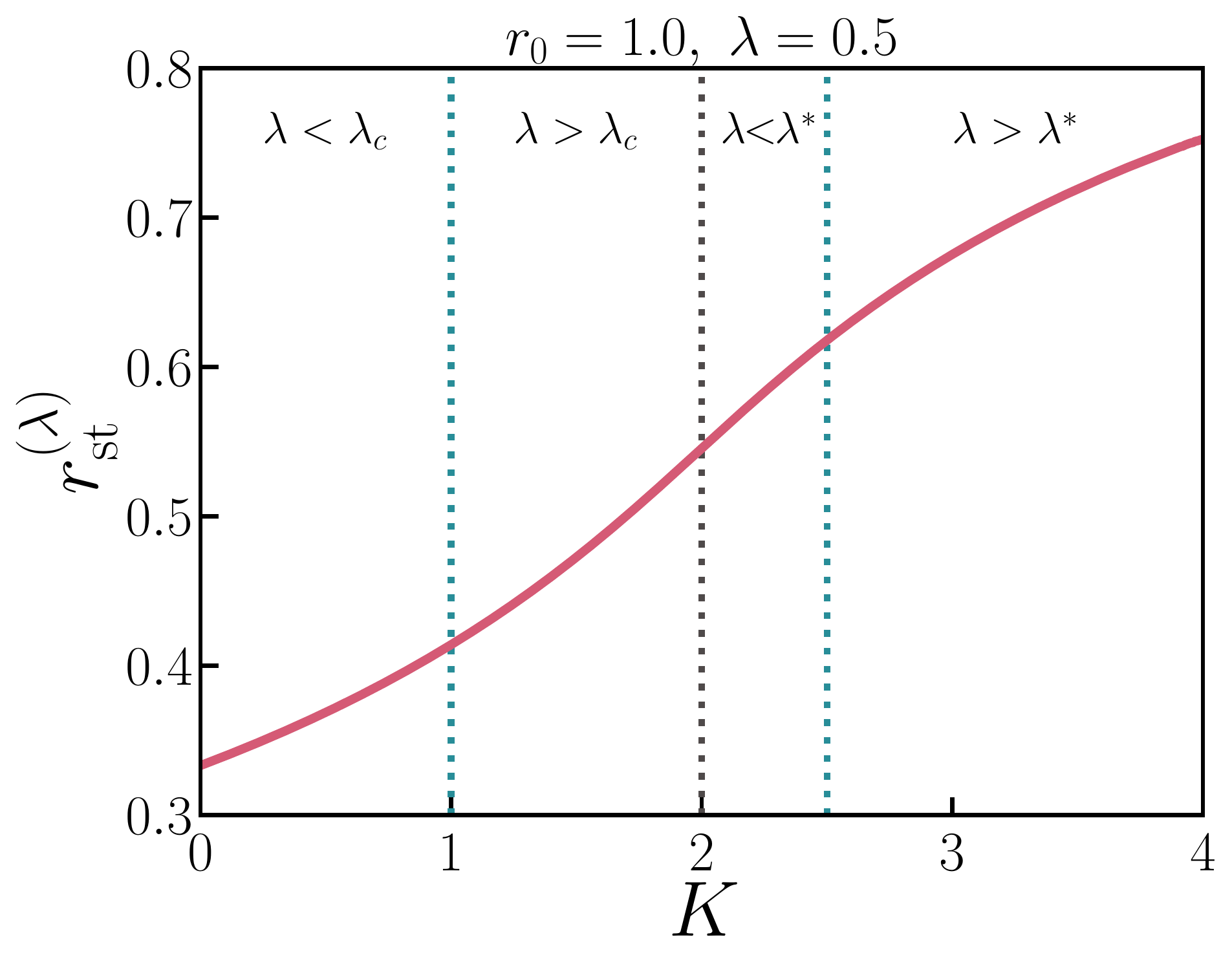}
	\caption{Behavior of the mean of the stationary-$r$ distribution, $P_{\rm st}(r)$, as a function of $K$, for the Lorentzian distribution of natural frequencies and for the choice of parameters: $r_0=1$, $\lambda = 0.5$ and $\sigma = 1.0$. The data are obtained by evaluating the integral $r_{\rm {st}}^{(\lambda)} = \int {\rm d}r ~ r P_{\rm st}(r)$ numerically using expressions of $P_{\rm st}(r)$ for the different parameter regimes as given in Subsec.~\ref{subsec:P_st_r0_gt_0}. The dashed lines at $K=1.0,~2.0$ and $2.5$ correspond to the thresholds $\lambda=\lambda_{c},~K=K_{c}$ and $\lambda = \lambda^\star$, respectively.}  
	\label{fig:Mean_r_st_lambda}
\end{figure}

One may note that, for any $\lambda >0$, $r_{\rm {st}}^{(\lambda)}$ is non-zero at any $K~\geq 0$. Thus, as soon as one introduces resetting in the Kuramoto dynamics, one will not encounter any phase transition between an unsynchronized phase and a synchronized phase while crossing any of the threshold lines corresponding to $\lambda_c$, $K_c$ and $\lambda^*$ in the phase diagram. The phase transition (bifurcation) occurs only at $\lambda =0$, i.e.,  under the bare Kuramoto dynamics. Referring to Fig.~\ref{fig:phase-diagram}, we have $r_{\rm {st}}^{(\lambda)} \neq 0$ for $\lambda >0$. Nevertheless, for $\lambda < \lambda_{c}$,  $P_{\rm st}(r)$ diverging as $r \to 0^+$ implies that in any typical realization of the dynamics, the system is unsynchronized. This motivated us to refer to the unshaded region in Fig.~\ref{fig:phase-diagram} as unsynchronized. Variation of $r_{\rm {st}}^{(\lambda)}$ with $K$ for representative values of $r_0$ and $\lambda$ is shown in Fig.~\ref{fig:Mean_r_st_lambda}. The data are obtained by evaluating the integral $r_{\rm {st}}^{(\lambda)} = \int {\rm d}r ~ r P_{\rm st}(r)$ numerically using expressions of $P_{\rm st}(r)$ for the different parameter regimes as given in Subsec.~\ref{subsec:P_st_r0_gt_0}.

The phase diagram bears strong resemblance with the one for the two-dimensional Ising model in presence of stochastic resetting studied in \cite{ising-resetting}.   Indeed,  just as stochastic resetting induces in our case a stationary ordered phase (the pseudo-synchronized phase) for values of the coupling strength $K$ satisfying $K<K_c$ for which the bare model does not exhibit any ordered phase (provided the resetting rate is larger than the threshold value $\lambda_c$), so is the situation with the Ising model.  In the latter,  the bare model does not show any ordered phase for temperatures larger than a critical temperature $T_c$. Yet, on introducing resetting, one observes in this very temperature range a stationary ordered phase, the pseudo-ferro phase, to be emerging so long as the resetting rate is larger than a threshold value.  There are other similarities as well:  in the Ising case, the threshold resetting rate separates the pseudo-ferro phase from a disordered phase, the paramagnetic phase, just in our case,  it separates the pseudo-synchronized phase from the unsynchronized phase. In fact, our nomenclature ``pseudo-synchronized phase'' is borrowed from Ref. ~\cite{ising-resetting}.  

However,  there are very important differences in the dynamical set-up of the Kuramoto and the Ising model that do not allow results in the latter to be applicable to the former.  The bare Ising model has a stochastic dynamics,  relaxes at long times to an equilibrium stationary state, and in fact has the attributes of a statistical system.  The bare Kuramoto model on the other hand has purely deterministic dynamics,  has a long-time stationary state that is not an equilibrium but is a nonequilibrium stationary state, and qualifies as a bona fide nonlinear dynamical system for which such overarching concepts of statistical physics as that the equilibrium state in a canonical ensemble setting is the one that minimizes the free energy do not apply.  There is one more crucial difference: the Ising model studied in Ref.~\cite{ising-resetting} does not have inherent quenched randomness,  unlike the Kuramoto model that involves quenched randomness in the form of the distributed natural frequencies of the Kuramoto oscillators.  It is remarkable that despite this quenched randomness, we are able to derive in this work exact analytical results for the phase diagram, thanks to the Ott-Antonsen ansatz,  which are in excellent agreement with numerical simulation results.

%%%%%%%%%%%%%%%%%%%%%%%%%%%%%%%%%%%%%%%%%%%%%%%%%%%%%%%%%%%%%%%%%%%%%%%%%%%%%%
\section{Gaussian frequency distribution: Numerical results}
\label{sec:Gaussian}
\begin{figure*}[ht!]
	\centering
	\includegraphics[scale=0.35]{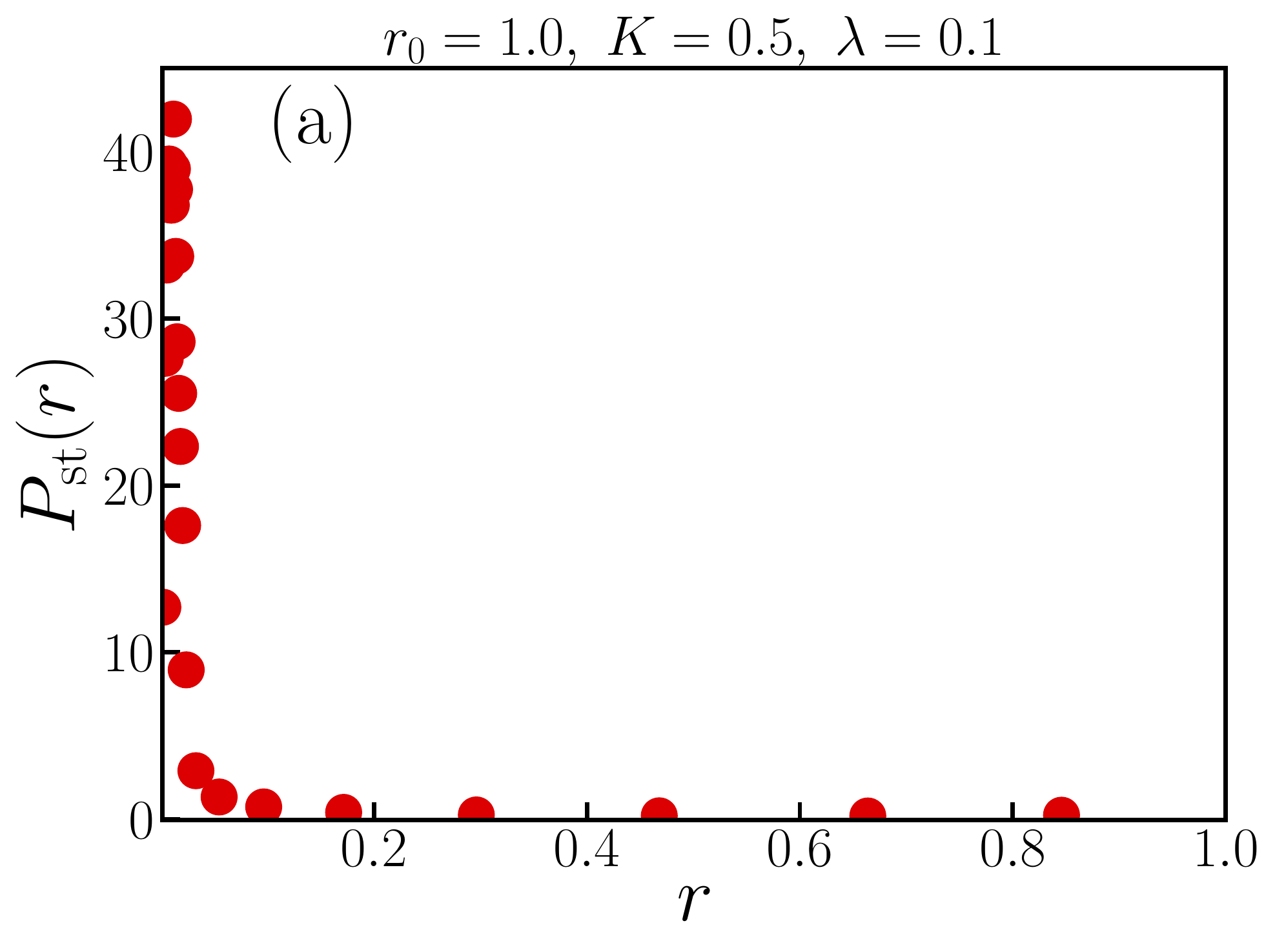}
	\includegraphics[scale=0.35]{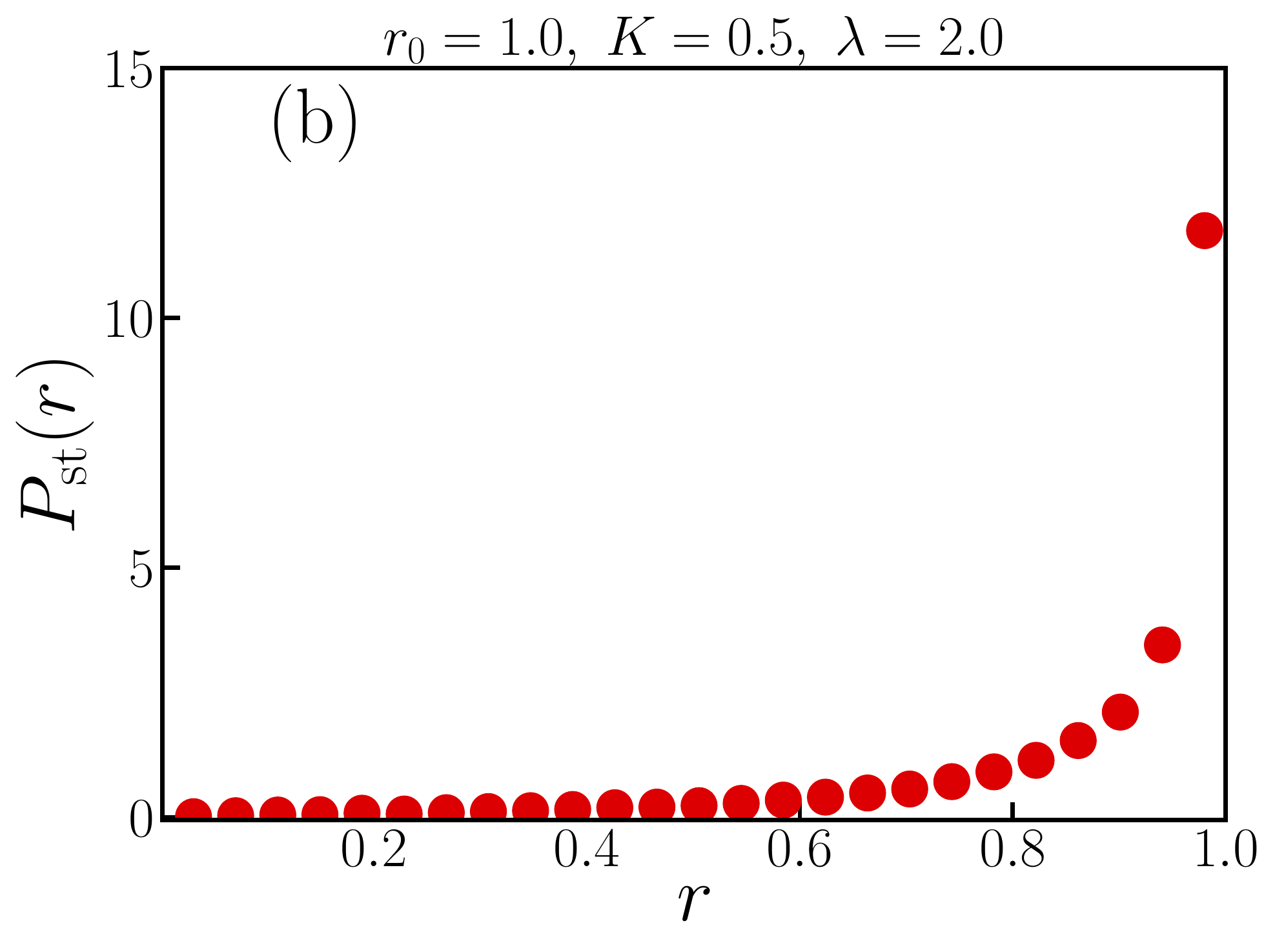}
	\includegraphics[scale=0.35]{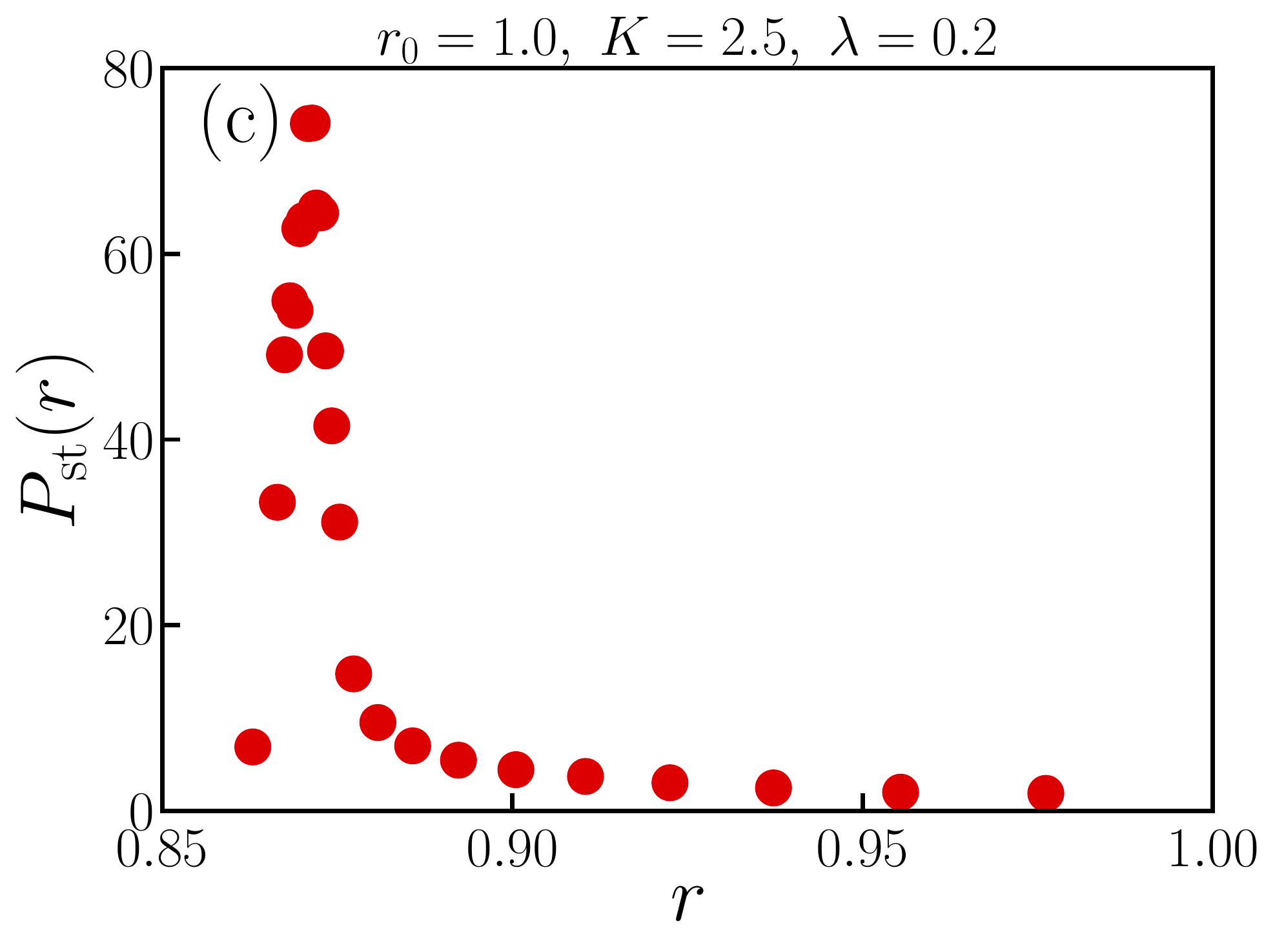}
	\includegraphics[scale=0.35]{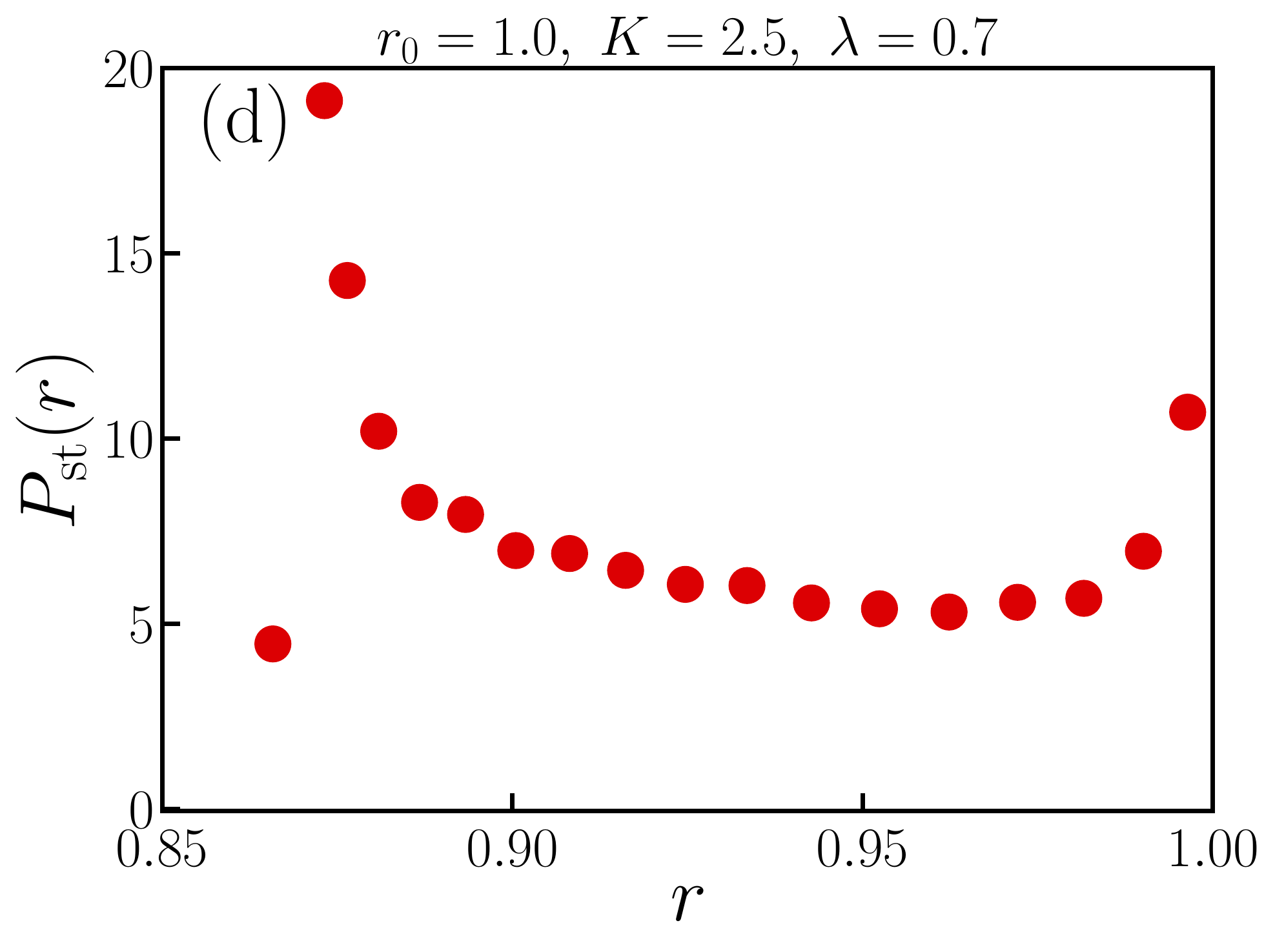}
	\includegraphics[scale=0.35]{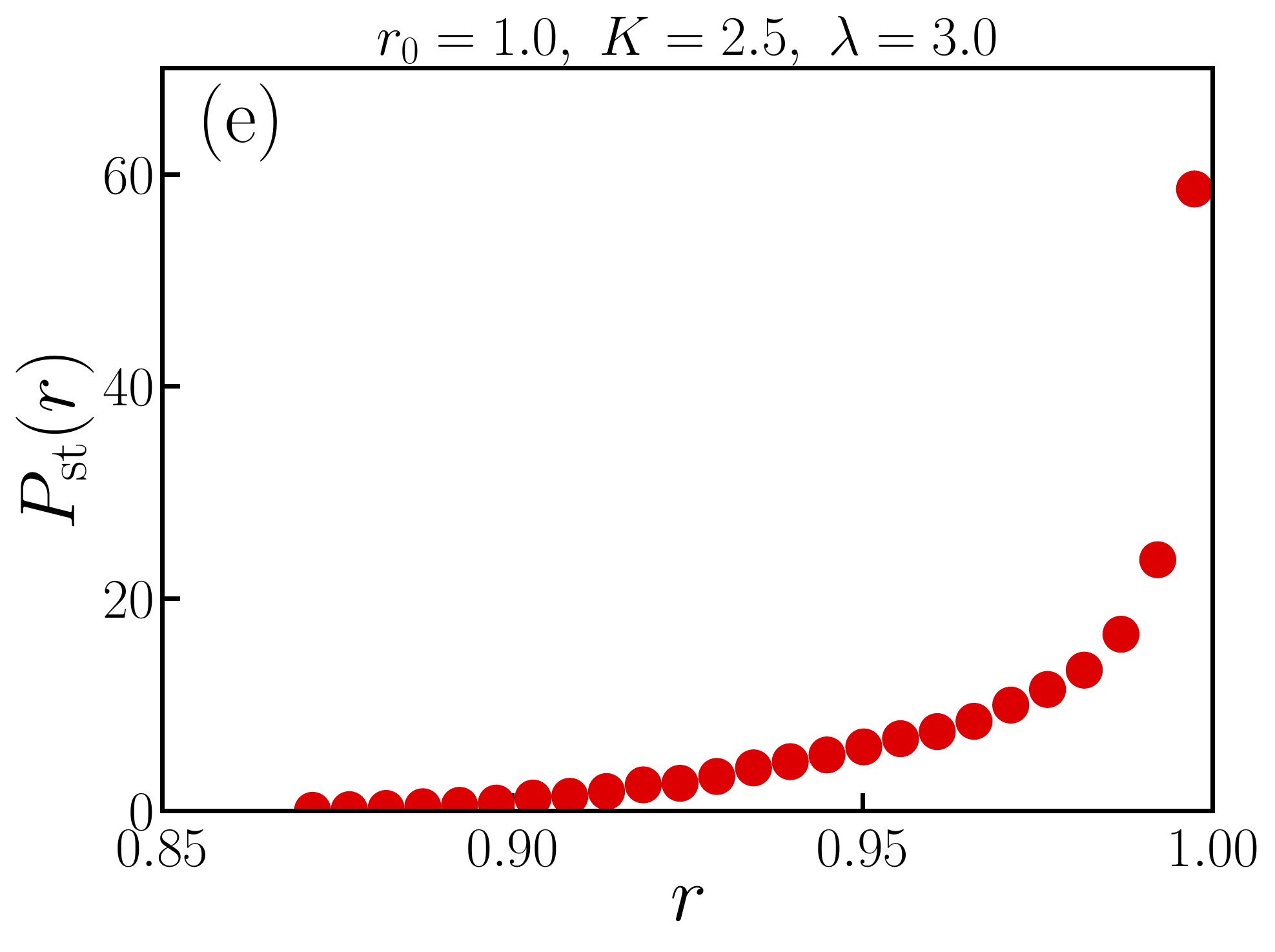}
	\caption{For the Gaussian distribution of the natural frequencies
				of the oscillators, Eq. ~(\ref{eq:Gaussian_lorentzian_g_omega}), with width $\sigma = 1.0$, shown is the stationary-state
				probability distribution $P_{\rm st}(r)$ based on $N$-body numerical simulation of the Kuramoto dynamics with resetting for number of oscillators $N = 10^4$ and reset parameter $r_0 = 1.0$.  For the chosen parameter values, one has $K_c=2\sqrt{2}/\sqrt{\pi}$.  In the figure,  panels (a) and (b) (respectively, panels (c), (d) and (e)) refer to values of $K$ smaller (respectively, larger) than $K_c$.   Panels (a), (c) and (d) (respectively, panels (b) and (e)) refer to small (respectively, large) values of the resetting rate $\lambda$.  The specific values of the $(K,\lambda)$ pair are $(0.5,0.1)$ for panel (a),  $(0.5,2.0)$ for panel (b),  $(2.5,0.2)$ for panel (c), $(2.5,0.7)$ for panel (d), and $(2.5,3.0)$ for panel (e). We have $r_{\rm st}=0$ for $K < K_c$, while for the chosen value of $K>K_c$,  the numerically estimated value of $r_{\rm st}$ is $r_{\rm st} \approx 0.87$, which is smaller than the chosen value of $r_0$.  Note that unlike for Lorentzian $g(\omega)$, one does not have for Gaussian $g(\omega)$ an analytical expression for $r_{\rm st}$ for $K> K_c$.  We note here that in conformity with the phase diagram in Fig.~\ref{fig:phase-diagram},  the stationary distribution $P_{\rm st}(r)$ is for $K<K_c$ defined for $r \in [0,r_0]$ and for $K>K_c$ defined for $r \in [r_{\rm st},r_0]$. }
	\label{fig:Gaussian-Pst}
\end{figure*}

In this section, we discuss results for the Gaussian distribution of the natural frequencies of the oscillators given by Eq.~(\ref{eq:Gaussian_lorentzian_g_omega}). Here, in the absence of a time evolution equation for the evolution of the order parameter $r(t)$ under the bare dynamics,  the analysis in presence of resetting that was adduced in Subsec.~\ref{sec:with-resetting} cannot be carried through. Consequently, we resort to numerical simulations of the dynamics, with the aim to investigate whether the general features of the phase diagram in Fig.~\ref{fig:phase-diagram} hold true also in this case.  Our numerical results are summarized in Fig.~\ref{fig:Gaussian-Pst}.  

Numerical simulation of the Kuramoto dynamics with resetting is performed with number of oscillators $N=10^4$ and a Gaussian distribution of the natural frequencies, Eq.~(\ref{eq:Gaussian_lorentzian_g_omega}), with width $\sigma=1.0$. For such a choice, the critical coupling strength of the bare Kuramoto model $K_c$ is given by $K_c= 2\sqrt{2}/\sqrt{\pi} \approx 1.5958$~\cite{Strogatz2000}. With no prior knowledge on the existence of any threshold $\lambda$ similar to $\lambda_{c}$ or $\lambda^*$  encountered for Lorentzian $g(\omega)$, we investigate the behavior of stationary-$r$ distribution for small and large $\lambda$ in both the regimes $K<K_c$ and $K>K_c$. In simulations, we choose a representative value of $r_0$, namely, $r_0=1.0$.

To analyse the behavior in the regime $K < K_c$, we choose $K=0.5~ (< K_c)$ and study the cases $\lambda = 0.1$ and $\lambda = 2.0$. We see that indeed, as shown in panel (a) and panel (b), one has the stationary distribution $P_{\rm st}(r)$ having near $r=0$ a peak and a vanishingly small value for small and large $\lambda$, respectively. This suggests in the limit $N \to \infty$ the corresponding behavior: $P_{\rm {st}}(r)$ diverging or $P_{\rm {st}}(r) \to 0$ as $r \to 0^+$, a behavior similar to the one with Lorentzian distribution, see Eq.~(\ref{eq:P_st_unsync_near_r_0}) and Fig.~\ref{fig:phase-diagram}. This observation suggests the existence of a threshold akin to $\lambda_c$ that we had reported in Eq.~(\ref{eq:lambdac}) for Lorentzian $g(\omega)$.

The situation is no different for $K>K_c$ as we change our distribution from Lorentzian to Gaussian. Here, we choose the coupling constant to be $K=2.5~ (> K_c)$ and take three values of $\lambda$, namely, $\lambda = 0.2, ~0.7$, and $\lambda = 3.0$. The choice of $K$ and $r_0$ implies that we have $r_0 > r_{\rm st}$. For the chosen smaller values of $\lambda$, as shown in panels (c) and (d), a peak appears in the distribution $P_{\rm {st}}(r)$ near $r=r_{\rm {st}}$, which may be attributed to finite-size effects. The appearance of a peak suggests diverging $P_{\rm {st}}(r)$ as $r \to r_{\rm {st}}^+$ in the limit $N \to \infty$. Additionally, panel (d) shows a dip in the stationary-$r$ distribution. On the other hand, when $\lambda$ is large, $P_{\rm {st}}(r)$ is vanishingly small near $r=r_{\rm {st}}$, see panel (e). This suggests $P_{\rm {st}}(r) \to 0$ as $r \to r_{\rm {st}}^+$ in the limit $N \to \infty$, a behavior similar to what was observed for the Lorentzian distribution, see Fig.~\ref{fig:phase-diagram}.

We have $r_{\rm st}=0$ for $K < K_c$, while for the chosen value of $K>K_c$, the numerically estimated value of $r_{\rm st}$ is $r_{\rm st} \approx 0.87$ for $N=10^4$, which is smaller than the chosen value of $r_0$.   Unlike for Lorentzian $g(\omega)$, one does not have for Gaussian $g(\omega)$ an analytical expression for $r_{\rm st}$ for $K> K_c$.  It may be noted from Fig.~\ref{fig:Gaussian-Pst} that in conformity with the phase diagram in Fig.~\ref{fig:phase-diagram},  the stationary distribution $P_{\rm st}(r)$ is for $K<K_c$ defined for $r \in [0,r_0]$ and for $K>K_c$ defined for $r \in [r_{\rm st},r_0]$. 

Note that we do not have an analytical form for $P_{\rm {st}}(r)$ for Gaussian $g(\omega)$. Yet,  what we have observed on the basis of our exact results for Lorentzian $g(\omega)$ as regards the behavior of $P_{\rm {st}}(r)$ as one approaches $r_{\rm {st}}$ in both the regimes $K <K_c$ and $K > K_c$ holds also in our numerical simulation results for Gaussian $g(\omega)$. Thus, our results for Gaussian $g(\omega)$ are all in conformity with the phase diagram in Fig.~\ref{fig:phase-diagram} for Lorentzian $g(\omega)$.  Moreover,  we see just as for the latter that for $K<K_c$ and large $\lambda$, one has a synchronized phase induced solely by the effects of stochastic resetting.  

We note that in contrast to the case of Lorentzian $g(\omega)$, we did not observe in our preliminary investigations for Gaussian $g(\omega)$ a $P_{\rm {st}}(r)$-profile with a peak similar to what is shown in Fig.~\ref{fig:phase-diagram}. In this regard, a more detailed investigation is required, which is planned for future work.

%%%%%%%%%%%%%%%%%%%%%%%%%%%%%%%%%%%%%%%%%%%%%%%%%%%%%%%%%%%%%%%%%%%%%%%%%%%%%%
\section{Conclusion}
\label{sec:conclusions}

In this work, we reported on how introducing the stochastic dynamics of resetting,  a paradigmatic framework of modern statistical physics,  modifies the phase diagram of a prototypical nonlinear dynamical system involving coupled oscillators.  In particular,  a remarkable revelation is that resetting is able to induce in the system a globally-synchronized phase in parameter regimes for which the bare model without resetting  does not support such a phase.  This fact paves the way to engineer ways to synchronize a bunch of interacting coupled oscillators that would not on their own be synchronized.  While the exact analytical results presented in this work were obtained for the case of Lorentzian distribution of natural frequencies of the oscillators,  an impending task would be to analytically characterize the numerical results that we reported here for the Gaussian distribution of frequencies. It would also be interesting to extend the analysis presented here to other coupled-oscillator systems allowing for a more general form of interaction than the Kuramoto model~\cite{Pikovskybook}. 

The resetting mechanism that we considered in this work is of global nature, in the sense that it involves all the oscillators undergoing resets simultaneously so that the entire system gets back to its initial state.  It would be interesting to consider the situation in which individual oscillators reset independently, i.e., the reset dynamics for every oscillator is an independent Poisson process with corresponding reset rate equal to, say, $\lambda_j$ for the $j$-th oscillator.  Similar set-ups have been considered in Ref.~\cite{miron2021diffusion} in the context of diffusing particles with excluded volume interactions and moving between the sites of a one-dimensional lattice, whereby the particles independently attempt to reset their position to the origin of the lattice.  The study unveiled interesting stationary-state behavior of the density profile of the particles, whose origin may be traced to an intricate interplay between interactions and resetting  in the dynamics. The Kuramoto model has the additional feature of quenched disorder arising from the distributed natural frequencies of the oscillators. We may anticipate this source of disorder to play a non-trivial role in dictating the stationary-state properties, the unveiling of which calls for a future study.

%%%%%%%%%%%%%%%%%%%%%%%%%%%%%%%%%%%%%%%%%%%%%%%%%%%%%%%%%%%%%
%

\appendix
\section{Details of numerical simulations} 
\label{app:Numerical_details}
\subsection*{Generating for a given realization $\{\omega_j\}$ the initial state $\{\theta_j(0)\}$, to which repeated resetting takes place:}
(a) For the choice $r_0=1$, the initial state corresponds to setting all the $\theta_j$s equal to a given value in $[0,2\pi)$.  (b) For the choice $r_0=0$, the initial state corresponds to distributing the $\theta_j$s uniformly on the interval $[0,2\pi)$.  For the case of the Lorentzian $g(\omega)$,  note that the values $r_0=0,1$ lie on the OA manifold.   (c) For the choice $0 < r_0 <1$,  we simulate the Kuramoto dynamics~(\ref{eq:eom0}) for given values of $N$ and $K>K_c$ and for the given realization of frequencies $\{\omega_j\}$, and while starting from an initial state with the $\theta_j$s uniformly distributed on $[0,2\pi)$.  To this end,  we integrate numerically the equations of motion~(\ref{eq:eom0}) by using a standard fourth-order Runge-Kutta algorithm with time step equal to $0.01$, and record the values of the $\theta_j$s at long times (typically, at times of order $10^4$) and the corresponding value of $r_0$.  The latter value will necessarily lie in the range $0 < r_0 <1$. The long-time values of the $\theta_j$ constitute the initial as well as the reset state for the given realization $\{\omega_j\}$ while implementing the Kuramoto dynamics with stochastic resetting.  The obtained long-time value of $r_0$ is what we use in the theoretical results to fit the data.  For the case of the Lorentzian $g(\omega)$, it is guaranteed by virtue of the attracting property of the OA manifold that the long-time state lies on the manifold, and thus, we are able to implement reset to a non-zero value of $r_0$ that lies on the OA manifold.  Note that in all cases (a),  (b), (c), the system during a reset event is always reset to the same state $\{\theta_j(0)\}$. 

\subsection*{Details of numerical simulation of the Kuramoto model with stochastic resetting:}
The various steps are as follows:
\begin{enumerate}
	\item Choose values of $N$,  $\lambda$, and the total simulation time ${\cal T}$. Choose also the frequency distribution $g(\omega)$. In our simulations, we have chosen ${\cal T} = 10^4$. We have checked that the dynamics attains stationary state well before $t \approx 0.25 {\cal T}$.
	\item Choose a quenched-disordered frequency realization $\{\omega_j\}$.
	\item Choose the initial state $\{\theta_j(0)\}$ (which is also the reset state) and the corresponding value of $r_0$.
	\item Starting from the initial state, we let the bare Kuramoto dynamics~(\ref{eq:eom0}) evolve for a random time $\tau$ sampled from the exponential distribution~(\ref{eq:exponential}). This step is implemented in numerics by integrating the equations of motion of the bare model with a fourth-order Runge-Kutta algorithm with time step equal to $0.01$.  At the end of the evolution, the state of the system is instantaneously reset to the initial state.
	\item Step 4 is repeated the required number of times so as to ensure that the total duration of evolution equals the chosen value ${\cal T}$.
	
	\item For a given realization of the dynamics, we record the values of $\{\theta_j(t)\}$ at various times $t \in [0.75 {\cal T}, {\cal T}]$, with successive recordings separated by a time gap larger than the mean time between two successive resets, i.e. $1/\lambda$. These $\{\theta_j(t)\}$s are used to obtain the value of $r$ at the corresponding times for the given realization of the dynamics.	
\end{enumerate}

Steps 2 -- 6 are repeated a number of times to obtain values of $r$ (typically, the sample size is of the order $10^5$), which are eventually used to construct the $r$-distribution reported in the main text. The $r$-distribution so constructed thus corresponds to an average over simulations for different frequency realizations. We use a uniform random-number generator to generate the Lorentzian $g(\omega)$ using standard procedure~\cite{Kalos}, while the Gaussian $g(\omega)$ is generated by using the standard Box-Muller algorithm~\cite{Kalos}. In passing, we remark that steps 2--4 are used to generate the time variation of $r$ in a given realization of the dynamics, as shown in Fig.~\ref{fig:r_evolution}.

\section*{Data Availability}
The data that support the findings of this study are available from the corresponding author upon reasonable request.

%%%%%%%%%%%%%%%%%%%%%%%%%%%%%%%%%%%%%%%%%%%%%%%%%%%%%%%%%%%%%%%%%%%%%%%%%%%%%%
\acknowledgments
MS thanks HPCE, IIT Madras for providing high performance computing facilities in AQUA cluster.  SG acknowledges support from the Science and Engineering Research Board (SERB), India under SERB-TARE scheme Grant No. TAR/2018/000023, SERB-MATRICS scheme Grant No. MTR/2019/000560, and SERB-CRG Scheme Grant No. CRG/2020/000596.  He also thanks ICTP – The Abdus Salam International Centre for Theoretical Physics, Trieste, Italy for support under its Regular Associateship scheme.

%%%%%%%%%%%%%%%%%%%%%%%%%%%%%%%%%%%%%%%%%%%%%%%%%%%%%%%%%%%%%%%%%%%%%%%%%%%%%%
%\bibliographystyle{aip}
\bibliography{References_mrinal}

%%%%%%%%%%%%%%%%%%%%%%%%%%%%%%%%%%%%%%%%%%%%%%%%%%%%%%%%%%%%%%%%%%%%%%%%%%%%%%
\end{document}